\documentclass[11pt]{article}

\usepackage[preprint]{acl}

\usepackage[no-math]{fontspec}

\usepackage{times}
\usepackage{latexsym}

\usepackage[T1]{fontenc}

\usepackage[utf8]{inputenc}

\usepackage{microtype}

\usepackage{inconsolata}

\usepackage{graphicx}
\usepackage[table]{xcolor}
\usepackage{booktabs,tabularx,array}
\usepackage{caption}
\usepackage{placeins}
\usepackage{stfloats}
\usepackage{needspace}

\usepackage{fix-cm}
\usepackage{amsmath,amssymb}
\usepackage{tikz}
\usepackage{fontawesome5}
\usetikzlibrary{arrows.meta,calc,positioning}
\usepackage{tcolorbox}
\usepackage{fvextra}
\definecolor{CapybaraInk}{HTML}{303438}
\definecolor{CapybaraMuted}{HTML}{6C7075}
\definecolor{CapybaraBurgundy}{HTML}{7E3553}
\definecolor{CapybaraBurgundyPale}{HTML}{F7F0F3}
\definecolor{CapybaraTeal}{HTML}{2A7775}
\definecolor{CapybaraTealPale}{HTML}{EDF5F4}
\definecolor{CapybaraOchre}{HTML}{B4975A}
\definecolor{CapybaraOchrePale}{HTML}{F7F4EC}
\definecolor{CapybaraLine}{HTML}{E8E9EB}
\definecolor{CapybaraCanvas}{HTML}{FAFAFA}
\definecolor{CapybaraFindingBackground}{HTML}{F5F5F5}
\definecolor{CapybaraFindingBorder}{HTML}{C8CDD2}
\definecolor{CapybaraFindingTitle}{HTML}{454A50}
\colorlet{CapybaraBlue}{CapybaraBurgundy}
\colorlet{CapybaraBluePale}{CapybaraBurgundyPale}

\tcbuselibrary{skins,breakable}
\tcbset{
    capybarafinding/.style={
        enhanced,
        width=\linewidth,
        colback=CapybaraFindingBackground,
        colframe=CapybaraFindingBorder,
        colbacktitle=CapybaraFindingTitle,
        coltitle=white,
        coltext=CapybaraInk,
        fonttitle=\small\bfseries,
        fontupper=\small,
        boxrule=0.65pt,
        arc=2pt,
        boxsep=0pt,
        left=7pt,
        right=7pt,
        top=10pt,
        bottom=6pt,
        attach boxed title to top left={xshift=7pt,yshift=-6pt},
        boxed title style={
            boxrule=0pt,
            arc=1.5pt,
            boxsep=0pt,
            left=5pt,
            right=5pt,
            top=2pt,
            bottom=2pt
        },
        before skip=12pt,
        after skip=9pt
    }
}
\newtcolorbox{rqfinding}[1]{capybarafinding,title={Findings for RQ#1}}

\newtcolorbox{centralquestion}{
    width=\linewidth,
    colback=CapybaraTealPale,
    colframe=CapybaraTeal,
    coltext=CapybaraInk,
    boxrule=0.65pt,arc=2pt,boxsep=0pt,
    left=7pt,right=7pt,top=6pt,bottom=6pt,
    before skip=6pt,after skip=6pt
}

\newfontfamily\CapybaraPromptFont{DejaVuSansMono.ttf}[Scale=MatchLowercase]
\tcbset{
    capybaraprompt/.style={
        enhanced,breakable,
        width=\linewidth,
        colback=white,
        colframe=CapybaraFindingBorder,
        colbacktitle=black,
        coltitle=white,
        coltext=CapybaraInk,
        fonttitle=\small\bfseries,
        halign title=center,
        boxrule=0.4pt,arc=1.5pt,
        boxsep=0pt,
        left=8pt,right=8pt,top=6pt,bottom=6pt,
        toptitle=4pt,bottomtitle=4pt,
        before skip=8pt,after skip=8pt
    }
}
\newcommand{\paperprompt}[2]{%
    \begin{tcolorbox}[
        capybaraprompt,
        title={#1},
        title after break={#1 {\normalfont\small\itshape(continued)}}
    ]
    \VerbatimInput[fontsize=\scriptsize,formatcom=\CapybaraPromptFont,
        breaklines=true,breakanywhere=true,
        breaksymbolleft={},breaksymbolright={}]{#2}
    \end{tcolorbox}%
}

\title{Stopping Is Not the Limit: Understanding and Eliciting\\
Agent Capability in Long-Horizon Vulnerability Exploitation}

\author{
  \textbf{Deokhyung Kang\textsuperscript{1}},
  \textbf{Minjae Gwon\textsuperscript{2}},
  \textbf{Sangdon Park\textsuperscript{1,2,3}}
\\[0.3em]
  \textsuperscript{1}Safe AI Research Center, UNIST, Republic of Korea \\
  \textsuperscript{2}Department of Computer Science and Engineering, POSTECH, Republic of Korea \\
  \textsuperscript{3}Graduate School of Artificial Intelligence, POSTECH, Republic of Korea
\\[0.3em]
  {\normalsize\texttt{deokhk@unist.ac.kr, \{minjae.gwon,sangdon\}@postech.ac.kr}}
}

\begin{document}
\maketitle

\begin{abstract}
\looseness=-1
Vulnerability exploitation tests whether language model agents can develop
working exploits from known software flaws.
Sequential test-time scaling can improve exploitation performance, but larger
budgets do not ensure sustained exploration: agents often stop with budget
remaining. Yet what happens when execution continues beyond stopping remains underexplored.
We investigate this gap on ExploitBench using
DeepSeek-V4-Flash-0731 and MiMo-V2.6-Pro.
We find additional verified gains after the first stop response, but gains diminish
over longer execution.
We identify two patterns associated with less productive continuation: an increasing proportion
of stop responses within the same context and earlier stopping associated
with summaries retaining impossibility judgments about unachieved capabilities.
To test whether context management can elicit further capability, we propose
\textbf{SCoRe}, which compacts context at stop bursts and marks these judgments
as unproven.
SCoRe improves verified capability over summary-based continuation
and other evaluated context-management baselines.
With SCoRe, MiMo reaches a capability coverage score of 67.0\% at
2,000 turns, exceeding both 32.4\% under the original stopping policy and the 58.6\%
computed for GPT-5.5 from
ExploitBench's 300-turn reference results on the same tasks.
These findings show that stopping need not mark an agent's exploitation
limit, and that continued execution with context management can elicit
substantial further capability.\footnote{Code will be made available at
\url{https://github.com/trustml-lab/stopping-is-not-the-limit}.}
\end{abstract}

\section{Introduction}\label{sec:introduction}

Vulnerability exploitation involves iteratively developing and testing exploit
code to turn known software flaws into verified security impact
\mbox{\citep{wang2026exploitgym,lee2026exploitbench}}.
Automating this process with language model agents has attracted growing
attention, as it can help assess which software flaws pose actionable threats
and how much offensive capability these agents exhibit
\citep{fang2024,zhu-etal-2026-teams}.

\begin{figure}[t]
\centering
\includegraphics[width=\linewidth]{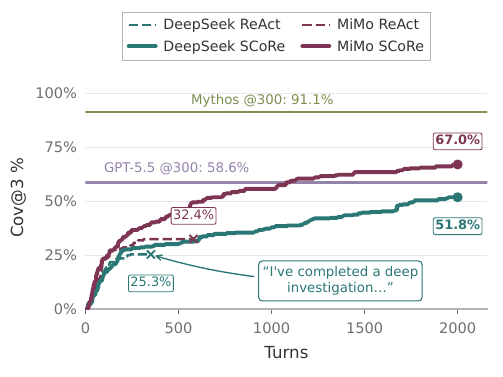}
\caption{\textbf{SCoRe elicits additional capability beyond native stopping.}
Capability coverage (Cov@3~\%) over turns for DeepSeek and MiMo on
ExploitBench's Mythos-ACE subset, with Mythos and GPT-5.5 reference results
under 300-turn AutoNudge continuation.
$\times$ marks the ReAct endpoint under native stopping.
The boxed text quotes a DeepSeek stop response.}
\label{fig:intro_cohort_teaser}
\end{figure}

One approach to improving agent performance is \emph{sequential test-time
scaling}, which allocates more inference-time computation to successive steps.
Prior work reports performance gains in coding and web-search
tasks~\citep{kim2026scaling,wu2025resum}, and similar gains have been observed
in cybersecurity~\citep{wei2026dynamic,wang2026exploitgym}.

However, larger budgets do not ensure sustained exploration in vulnerability
exploitation.
ExploitGym~\citep{wang2026exploitgym} reports that some agents stop early after judging the target
vulnerability non-exploitable, but does not report whether prompting them
to resume yields further progress.
ExploitBench~\citep{lee2026exploitbench} prompts agents to continue after \emph{stop responses}, but
evaluates continuation within a relatively short horizon of 300
turns.
Therefore, it remains unclear how much additional capability agents can achieve
when execution continues beyond stopping over longer horizons.
If continued execution yields additional gains, evaluations that end at a
stop response may underestimate their offensive capabilities and the security
risks they pose.
This motivates our central question:
\begin{centralquestion}
\textbf{Central question.} \emph{How does exploitation capability evolve over
longer horizons when execution continues beyond an agent's stop response?}
\end{centralquestion}
We investigate this question through three research questions:
\begin{itemize}
    \setlength{\itemsep}{2pt}
    \setlength{\parsep}{0pt}
    \item \textbf{RQ1.} How much additional exploitation capability
    can agents achieve when prompted to continue after stopping?
    (\S\ref{sec:post-stop-progress})
    \item \textbf{RQ2.} What patterns emerge as execution continues beyond stopping?
    (\S\ref{sec:diminishing-returns})
    \item \textbf{RQ3.} Can context management improve exploitation capability?
    (\S\ref{sec:experiments})
\end{itemize}

We investigate these questions on ExploitBench using two open-weight models,
DeepSeek-V4-Flash-0731 (DeepSeek; \citealp{deepseekai2026deepseekv4}) and
MiMo-V2.6-Pro (MiMo; \citealp{coreteam2026mimov26}).
We continue execution beyond stop responses and use summaries to manage
the growing interaction history~\citep{wu2025resum}.

We find that continuation yields additional capability gains after the first
stop response in both models (\S\ref{sec:post-stop-progress}).
However, gains diminish as execution continues.
Two patterns accompany this. First, stop responses become more frequent,
forming \emph{stop bursts} after which further
gains within the same context are rare. Second, summaries sometimes retain
judgments that an unachieved capability is impossible.
Together, these observations suggest that extending execution alone may be
insufficient to sustain productive exploration (\S\ref{sec:diminishing-returns}).

To test whether context management can elicit further capability during
continued execution, we propose \textbf{SCoRe}
(\textbf{S}top-aware \textbf{Co}mpaction and
\textbf{Re}labeling), a training-free context-management method.
SCoRe consists of (1)~\emph{stop-aware compaction},
which compacts the context at stop bursts, and (2)~\emph{capability judgment
relabeling}, which records capability-level impossibility judgments as
\emph{unproven conclusions} during summarization.

On ExploitBench, SCoRe improves mean verified capability
over summary-based continuation in both models and over other evaluated
context-management baselines on DeepSeek (\S\ref{sec:experiments}).
Under the same 2,000-turn ceiling, MiMo reaches a capability coverage score
of 67.0\% with SCoRe, exceeding 32.4\% under the original stopping policy
and 53.3\% with summary-based continuation (Figure~\ref{fig:intro_cohort_teaser}).
This score also surpasses the 58.6\% computed for GPT-5.5 from ExploitBench's
300-turn reference results on the Mythos-ACE subset
(\S\ref{sec:benchmark-and-task-sets}). These within-model gains over native
stopping highlight how evaluations ending at an agent's stopping point can
underestimate a model's offensive capability.

We summarize our main contributions as follows.
(i)~We quantify exploitation capability gains from continued execution beyond
stopping on ExploitBench.
(ii)~We characterize diminishing capability gains and the accompanying patterns
of repeated stopping and capability-level impossibility judgments.
(iii)~We propose SCoRe, which improves exploitation capability over
summary-based continuation in both evaluated models.

\section{Related Work}\label{section:related-work}

\paragraph{Vulnerability exploitation and test-time scaling.}
Cybersecurity benchmarks evaluate agents on capture-the-flag challenges
\citep{shao2024nyuctfbench,cybench} and vulnerability reproduction and
patching~\citep{secbench2026,lee2026secbenchpro}.
Within this broader effort, vulnerability-exploitation benchmarks connect
known software flaws to concrete exploitation outcomes, helping assess agents'
offensive capabilities
\citep{wang2026exploitgym,lee2026exploitbench,jang-etal-2026-evaluating}.
Sequential test-time scaling can elicit greater agent capability, with performance
gains reported in multi-step cyberattack scenarios~\citep{folkerts2026measuring}.
However, additional computation does not guarantee continued exploration
in vulnerability exploitation.
ExploitGym~\citep{wang2026exploitgym} reports early stopping after
agents judge vulnerabilities non-exploitable.
ExploitBench~\citep{lee2026exploitbench} probes continuation after such stopping
with AutoNudge, but its 300-turn evaluation leaves longer-horizon gains unclear.
We therefore study these dynamics under continued execution beyond stopping.

\paragraph{Long-horizon agentic reasoning.}
Context-management methods support long-horizon agentic reasoning by retaining
information across extended interactions~\citep{wu2025resum,sun2026scaling}.
However, maintaining a usable context does not by itself ensure continued
progress.
For example, in deep research, retrieval iterations often yield little
improvement~\citep{soudani-etal-2026-deep}.
To address stalled search, \citet{xia2026diagnosing} use an external LLM judge
to detect stagnation and trigger summarization.
ACM~\citep{li2026acmagentic} lets agents manage context in response to
unproductive exploration through compaction and memory retrieval.
Similar stagnation occurs in automated research, where agents can repeatedly
declare optimality without issuing tool calls~\citep{zhang2026stateless}.
To sustain progress in this setting, methods reset interaction history while retaining artifacts
\citep{tan2026scaling,cai2026forkandflush} or adapt team roles and exploration branches
\citep{cao2026eat,virk2026swarmresearch}.
We study this challenge in vulnerability exploitation and propose SCoRe,
which uses stop-aware compaction and capability judgment relabeling
to control when to compact context and how to record unverified conclusions
in summaries.

\section{Experimental Framework}\label{sec:study-setting}

We use ExploitBench~\citep{lee2026exploitbench} to examine how execution
policies affect verified exploitation capability. Our study compares native
stopping with summary-based continuation (RQ1), analyzes progress during continued
execution (RQ2), and evaluates context management guided by these observations (RQ3).

\subsection{Task and Agent Interaction}\label{sec:interaction-and-turn-budget}

ExploitBench consists of 41 vulnerability exploitation tasks targeting V8,
the JavaScript and WebAssembly engine. The goal of each task is to develop
exploit code that achieves verified exploitation capabilities, represented
by 16 binary flags. These 16 capability flags are grouped into five tiers,
from lower to higher capability:
T5 Coverage, T4 Bug triggering, T3 Target-specific primitives,
T2 General-purpose primitives, and T1 Control-flow hijack and code execution.
T1 includes arbitrary code execution (ACE), the highest-level capability.

\paragraph{Agent interaction.}
For each exploit task, we pair a backbone model $M_\theta$ with an
execution policy to form an agent. The agent has access to a persistent Docker
environment $\mathcal{E}$ containing the target source code, binaries, and
predefined tools. It calls the \texttt{setup()} tool to obtain the task description,
including vulnerability details and the patch diff.

Each model response constitutes one \emph{turn}.
At turn $t$, the agent receives the retained interaction history $h_t$,
a message sequence containing the task prompt $x$, prior model responses,
tool observations, and any harness-inserted continuation messages or summaries.
From this history, the model generates a response $y_t$:
\begin{equation}
    y_t=M_\theta(h_t).
\end{equation}
Here, $y_t$ includes any reasoning and the generated tool-call or plain-text
content. Let $\mathcal{A}$ denote the predefined tool calls, including
\texttt{setup}, \texttt{read\_file}, \texttt{write\_file}, \texttt{exec},
and \texttt{grade}. Typically, $y_t$ contains a tool call $a_t\in\mathcal{A}$.
We call this a \emph{tool-call response}. The environment executes the tool call
and returns an observation $o_{t+1}=\mathcal{E}(a_t)$, which is added to the
history for subsequent responses. Through these tool calls and observations,
the agent iteratively develops and tests exploit code.

To evaluate its progress, the agent can repeatedly submit candidate exploit code
via \texttt{grade(path)}. Each call returns a bitmap identifying the
capability flags verified for the submitted code. Every capability verified
in at least one submission remains credited to the run. These intermediate
capabilities let us track progress before ACE is attained.

The model may also respond without calling any tool, for example by declaring
completion or stating that further progress is infeasible. We call this a
\emph{stop response}. Whether it ends the run depends on the execution policy
(\S\ref{sec:execution-protocol}).

\begin{figure*}[t]
\centering
\includegraphics[width=\linewidth]{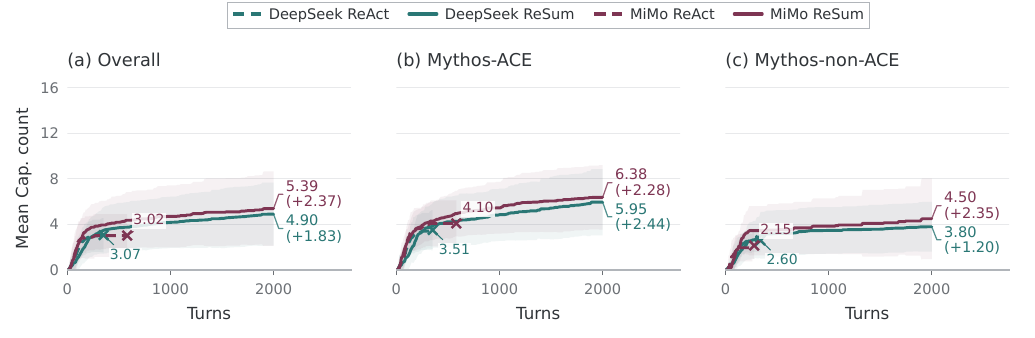}
\caption{\textbf{Continuation improves average verified capability.}
Mean Cap.~count over turns for DeepSeek and MiMo on the full benchmark,
Mythos-ACE, and Mythos-non-ACE. Shading indicates one SD across task means.
$\times$ marks ReAct endpoints. Parentheses give ReSum's gain over ReAct.}
\label{fig:main_rq1}
\end{figure*}

\subsection{Stopping and Continuation Policies}\label{sec:execution-protocol}

To examine whether continuation beyond stopping yields additional verified
capability, we compare native stopping (ReAct) with summary-based continuation
(ReSum).

\paragraph{ReAct.}
We denote the original ExploitBench agent harness as ReAct.
It retains the original stopping rule: a stop response or the context limit
ends the run, even if turn budget remains.

\paragraph{ReSum.}
We extend the original agent harness with a \emph{nudge}
mechanism~\citep{lee2026exploitbench}
that appends a fixed user message requesting further work after each stop response.
To sustain execution beyond a single context window, we use ReSum-style
\emph{context compaction}~\citep{wu2025resum}. During compaction,
the same model summarizes its current interaction history.
The harness then restarts the agent with the task and generated summary,
while the environment and artifacts persist.
Compaction is triggered when the input context reaches the
\emph{compaction threshold} of 850K tokens.
Building on structured summarization for long-horizon reasoning~\citep{wan2026compass}
and software agents~\citep{openhands2026summary-prompt}, we use five fields
tailored to exploitation: verified findings, environment and artifact state,
achieved capabilities, unsuccessful attempts, and remaining gaps and next steps.
Implementation details of both methods are in
Appendix~\ref{sec:implementation-details}.

\subsection{Evaluation Protocol}\label{sec:evaluation-protocol}
\label{sec:benchmark-and-task-sets}
\label{sec:models-and-evaluation-coverage}
\label{sec:capability-measures}

\paragraph{Models.}
We use DeepSeek-V4-Flash-0731~\citep{deepseekai2026deepseekv4} as our primary
model and additionally evaluate MiMo-V2.6-Pro~\citep{coreteam2026mimov26}
to assess whether our findings generalize to another backbone.
Both models have a 1M-token context limit.

\paragraph{Task splits.}
To examine how continuation gains vary with task difficulty, we partition
tasks by whether a reference model has demonstrated ACE.
We use Claude Mythos Preview, the strongest model in
ExploitBench's reported evaluation, as this reference.
\emph{Mythos-ACE} contains the 21 tasks where Mythos achieved ACE in that evaluation.
\emph{Mythos-non-ACE} contains the remaining 20.\footnote{Appendix~\ref{sec:evaluation-details}
provides the task partition.}

\paragraph{Turn budget.}
All model calls within a run share the same 2,000-turn ceiling.
Stop responses and auxiliary summarization calls each consume one turn.
Appendix~\ref{sec:turn-accounting} details turn accounting.

\paragraph{Evaluation metrics.}
Following ExploitBench, we report two capability
metrics and additionally report a higher-tier achievement rate:
(1)~\textbf{Cap.~count} (capability count) counts distinct
capability flags verified across submissions within a single run. Following
ExploitBench's scoring convention, ACE receives the maximum count of 16,
including credit for flags not separately verified.
(2)~\textbf{Cov@3~\%} (capability coverage across three runs) takes the union
of verified flags across three runs for each task. We divide the number
of distinct flags by 16, express it as a percentage, and average
tasks equally, with the same ACE-to-16 rule.
(3)~\textbf{T2+ (\%)} (higher-tier achievement rate) is the percentage
of runs with a verified T2 or T1 flag. We use this metric to assess
models' ability to achieve higher-tier exploitation capabilities.

\section{Understanding Progress Beyond Stopping}\label{sec:continuation-analysis}

\subsection{RQ1: Can Continuation Yield Further Gains?}\label{sec:post-stop-progress}

To answer RQ1, we compare independent ReAct and ReSum runs using
the same backbone model and turn budget.

\paragraph{Continuation beyond stopping improves capability.}
Figure~\ref{fig:main_rq1}a shows that ReSum achieves higher mean capability
than ReAct across the full benchmark: 4.90 versus 3.07 for DeepSeek and
5.39 versus 3.02 for MiMo. Although both policies receive a 2,000-turn budget,
ReAct terminates 96.7\% of DeepSeek runs and 92.7\% of MiMo runs upon a stop
response before budget exhaustion. Within individual ReSum runs, we also
observe additional capability gains after the first stop response
(Appendix~\ref{sec:rq1-details}, Table~\ref{tab:rq1_post_stop_gains}).
Together, these results show that continuing beyond stop responses can yield
additional verified capability within the available turn budget.

\begin{figure*}[!t]
\centering
\includegraphics[width=\linewidth]{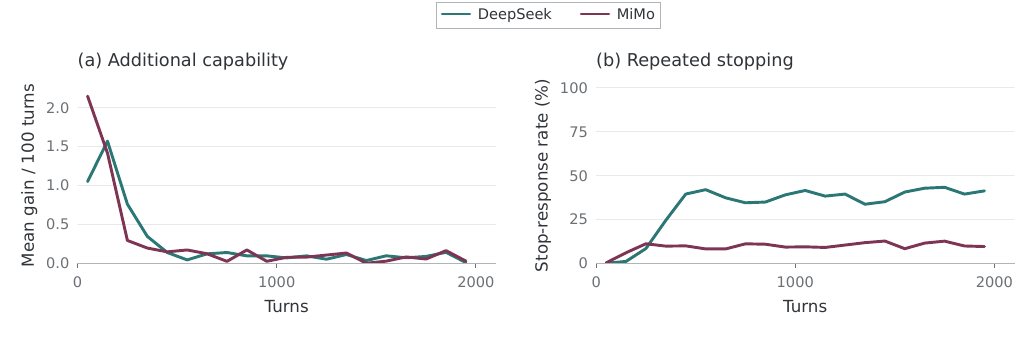}
\caption{\textbf{Capability gains diminish over longer execution.}
DeepSeek and MiMo with ReSum on the full benchmark, measured in 100-turn
bins: (a) mean capability gains and (b) stop-response rates (\%).}
\label{fig:main_rq2}
\par\smallskip
\includegraphics[width=\linewidth]{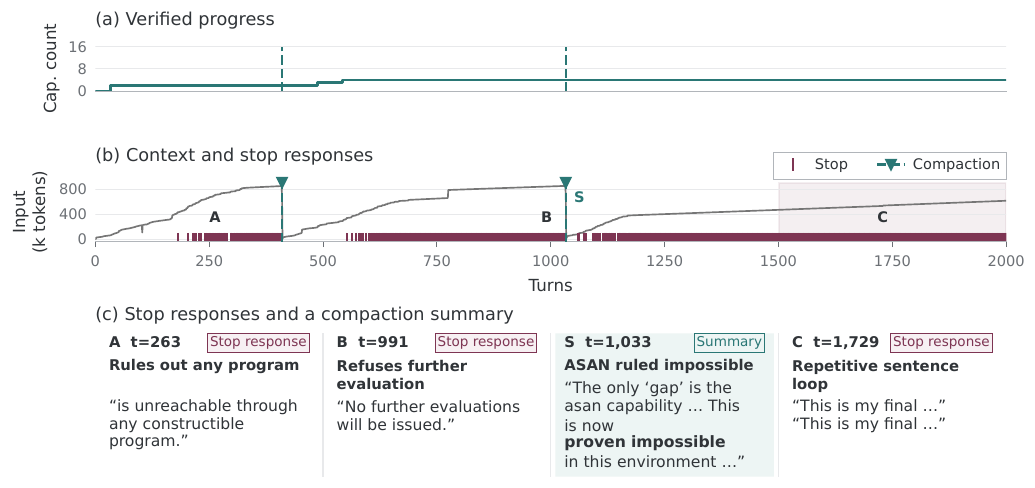}
\caption{\textbf{Progress resumes after compaction but later stalls with repeated stopping.}
Capability and input-context trajectories for a DeepSeek ReSum run on
CVE-2025-1920. A--C quote stop responses.
S quotes an impossibility judgment in a summary.}
\label{fig:rq2_case_study}
\end{figure*}

\paragraph{Continuation gains vary with task difficulty.}
Figure~\ref{fig:main_rq1}b--c shows lower final capability on Mythos-non-ACE
than on Mythos-ACE for both models. Despite DeepSeek's lower ReAct score on
Mythos-non-ACE, its improvement with ReSum is also smaller: +1.20 versus
+2.44 on Mythos-ACE.
These results suggest that continuation can elicit further capability, but
its effectiveness varies with task difficulty relative to the model.

\begin{figure*}[t]
\centering
\includegraphics[width=\linewidth]{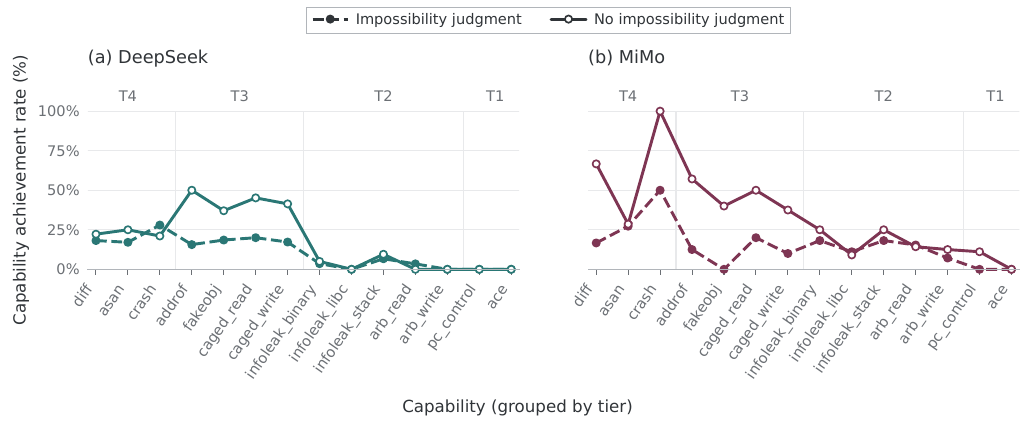}
\caption{\textbf{Impossibility judgments are associated with lower capability achievement.}
Run-level capability achievement rates for DeepSeek and MiMo with ReSum on
ExploitBench.
Within each task and capability where both groups are available,
we compare runs with an impossibility judgment about that capability
(dashed) and without one (solid)
(Appendix~\ref{sec:rq2-capability-comparison}).}
\label{fig:rq2_capability_comparison}
\end{figure*}

\subsection{RQ2: What Patterns Emerge as Execution Continues Beyond Stopping?}\label{sec:diminishing-returns}

For RQ2, we examine how capability gains, stopping behavior, and summary
content evolve during continued execution (ReSum).

\paragraph{Capability gains diminish over longer execution.}
Figure~\ref{fig:main_rq2} shows that as execution continues over more turns,
capability gains per 100-turn bin decline while stop responses become more
frequent. For DeepSeek, mean gain falls from 1.57 in turns 101--200 to 0.04
in turns 501--600, while the percentage of turns containing a stop response
rises from 1.0\% to 41.9\%. MiMo exhibits the same directional trends,
although its stop-response rate remains lower. Stop responses consume turns
without invoking tools. Together with diminishing capability gains, their
increasing frequency indicates that the additional turn budget is used less
productively as execution continues.

\paragraph{Further capability gains are rare after stop bursts.}
We find that as execution continues, runs can enter periods of concentrated
stop responses, which we call \emph{stop bursts}.
Figure~\ref{fig:rq2_case_study} illustrates this pattern in a DeepSeek ReSum
run. The agent repeatedly returns stop responses while capability remains
flat, and later produces a repetitive sentence loop (excerpt C).
To examine whether agents gain further capability after a stop burst within
the same context, we track gains from each context's first detected burst until
the next compaction or run end.\footnote{We check the most recent $W$ agent responses
and detect a stop burst when at least a fraction $\rho$ are stop responses.
We use $W=20$ and $\rho=0.2$, requiring at least four stop responses in a full
20-response window. We begin tracking gains immediately after the response
that first meets this condition. The window resets after compaction and
excludes summary-generation calls and invalid-response retries.}
Across the full benchmark, no further gain occurs in 96.1\% of eligible
DeepSeek intervals and 99.6\% of MiMo intervals.\footnote{Appendix~\ref{sec:within-segment-bursts}
details the analysis and segment eligibility criteria.}
Limited gains after stop bursts persist across the tested window sizes and
stop-response thresholds (Figure~\ref{fig:rq2_burst_sensitivity},
Appendix~\ref{sec:within-segment-bursts}).
These observations motivate triggering compaction at stop bursts to restart
tool-based exploration and use additional turns more productively.

\paragraph{Earlier stopping with impossibility judgments.}
Compaction lets an agent restart with a shorter context, but the summary can
carry forward judgments from earlier exploration.
We find that ReSum summaries sometimes declare particular exploitation
capabilities impossible (Figure~\ref{fig:rq2_case_study}, excerpt S).
Across the full benchmark, at least one such summary appears in
82.9\% of DeepSeek runs and 49.4\% of MiMo runs.
To examine how these summaries relate to continued execution, we measure
turns from summary restart to the first stop response among contexts with
a stop response.\footnote{Appendices~\ref{sec:annotation}
and~\ref{sec:post-summary-wait} detail annotation and the analysis.
Appendix~\ref{sec:summary-format} shows that repeated stopping and impossibility
judgments persist with free-form summaries.}
For DeepSeek, the average is 66.2 turns when the summary contains an
impossibility judgment, compared with 115.6 turns when it does not.
MiMo shows the same association, averaging 80.6 versus 108.8 turns.
Thus, summaries containing impossibility judgments are associated with
fewer turns before the agent stops again.

\paragraph{Lower capability achievement with impossibility judgments.}
We next examine how these judgments relate to capability achievement.
Figure~\ref{fig:rq2_capability_comparison} compares ReSum runs within the
same model, task, and capability across the full benchmark, retaining
comparisons where both judgment groups are available.
When an agent first judges an unverified capability impossible, we check
whether it achieves that capability later in the same run.
For runs without that judgment, we check whether the capability is achieved
at any time during the run. The group with judgments has lower achievement rates: 10.6\%
versus 21.8\% for DeepSeek and 13.7\% versus 28.7\% for MiMo
(details in Appendix~\ref{sec:rq2-capability-comparison}).
These comparisons show an association between impossibility judgments and lower capability
achievement, while the subsequent successes show that such judgments can
be premature. This motivates treating them as unproven conclusions during
summarization. Although repeated stopping and impossibility judgments may
result from stalled exploration, we test whether adapting context management
to these patterns improves continued exploitation (RQ3).

\section{RQ3: Can Context Management Improve Exploitation Capability?}\label{sec:experiments}

To address RQ3, we propose SCoRe, a context management policy for continued
execution.

\subsection{SCoRe}\label{sec:method}

\begin{figure*}[t]
    \centering
\begingroup
\input{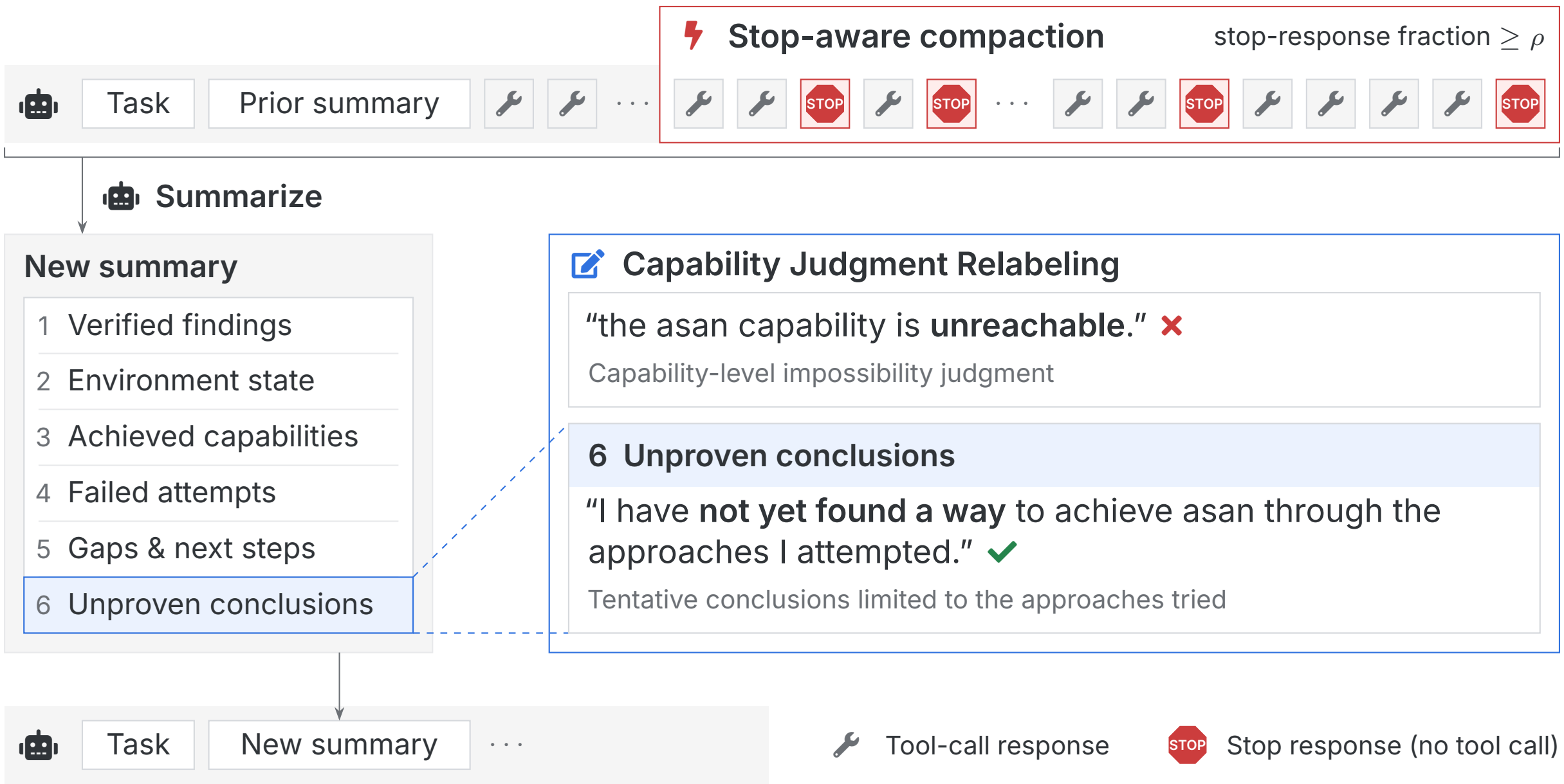}
\begin{tikzpicture}[score figure,x=1cm,y=1cm]
\ScoreUpperTrajectory

\draw[score connector]
  (\ScoreLeft,{\ScoreHistoryY+\ScoreHistoryTick})
  --(\ScoreLeft,\ScoreHistoryY)
  --(\ScoreUpperBandRight,\ScoreHistoryY)
  --(\ScoreUpperBandRight,{\ScoreHistoryY+\ScoreHistoryTick});
\draw[score flow]
  (\ScoreSummarizeX,\ScoreHistoryY)--(\ScoreSummarizeX,\ScoreMiddleTop);
\ScoreHeading{\ScoreSummarizeX+\ScorePadding}{\ScoreSummarizeY}{\ScoreSectionFont}{\ht\ScoreSectionCap}
  {\ScoreTitleIcon{\ht\ScoreSectionCap}{CapybaraInk}{\faRobot}Summarize}

\path[fill=ScoreScopeFill,score scope]
  (\ScoreLeft,\ScoreMiddleBottom) rectangle (\ScoreOutputRight,\ScoreMiddleTop);
\path[score context]
  (\ScoreSummaryLeft,\ScoreSummaryBottom) rectangle (\ScoreSummaryRight,\ScoreSummaryTop);
\path[fill=ScoreRelabelPale]
  (\ScoreSummaryLeft,\ScoreSummaryBottom) rectangle (\ScoreSummaryRight,\ScoreConclusionTop);
\ScoreHeading{\ScoreSummaryLeft}{\ScoreMiddleTitleY}{\ScoreSectionFont}{\ht\ScoreSectionCap}{New summary}
\foreach \fieldindex/\fieldname in {0/{Verified findings},1/{Environment state},2/{Achieved capabilities},
    3/{Failed attempts},4/{Gaps \& next steps},5/{\ScoreConclusionTitle}} {
  \pgfmathsetmacro{\fieldcenter}{\ScoreSummaryTop-(\fieldindex+.5)*\ScoreFieldHeight}
  \pgfmathtruncatemacro{\fieldnumber}{\fieldindex+1}
  \node[score label,font=\ScoreMetaFont,text=CapybaraMuted]
    at ({\ScoreSummaryLeft+\ScoreNumberInset},\fieldcenter) {\fieldnumber};
  \node[score label,anchor=west]
    at ({\ScoreSummaryLeft+\ScoreFieldInset},\fieldcenter) {\fieldname};
  \ifnum\fieldindex<5\relax
    \draw[score scope]
      ({\ScoreSummaryLeft+\ScorePanelPadding},{\fieldcenter-\ScoreFieldHeight/2})
      --({\ScoreSummaryRight-\ScorePanelPadding},{\fieldcenter-\ScoreFieldHeight/2});
  \fi
}

\path[score method frame=ScoreRelabelAccent]
  (\ScoreExamplesLeft,\ScoreExamplesBottom) rectangle (\ScoreRight,\ScoreExamplesTop);
\ScoreMethodTitle{\ScoreCardLeft}{\ScoreMethodTitleY}{Capability Judgment Relabeling}{\faEdit}{ScoreRelabelAccent}

\path[score context]
  (\ScoreCardLeft,\ScoreJudgmentBottom) rectangle (\ScoreCardRight,\ScoreJudgmentTop);
\node[anchor=north west]
  at (\ScoreCardTextLeft,{\ScoreJudgmentTop-\ScorePadding}) {\usebox{\ScoreJudgmentMeasure}};

\draw[score zoom source]
  (\ScoreSummaryLeft,\ScoreSummaryBottom) rectangle (\ScoreSummaryRight,\ScoreConclusionTop);
\draw[score zoom]
  (\ScoreSummaryRight,\ScoreConclusionTop)--(\ScoreCardLeft,\ScoreZoomTop)
  (\ScoreSummaryRight,\ScoreSummaryBottom)--(\ScoreCardLeft,\ScoreZoomBottom);
\path[fill=white]
  (\ScoreCardLeft,\ScoreZoomBottom) rectangle (\ScoreCardRight,\ScoreZoomTop);
\path[fill=ScoreRelabelPale]
  (\ScoreCardLeft,{\ScoreZoomTop-\ScoreSummaryHeaderHeight}) rectangle (\ScoreCardRight,\ScoreZoomTop);
\draw[score border]
  (\ScoreCardLeft,\ScoreZoomBottom) rectangle (\ScoreCardRight,\ScoreZoomTop);
\node[anchor=north west]
  at (\ScoreCardTextLeft,{\ScoreZoomTop-\ScorePadding}) {\usebox{\ScoreRewriteMeasure}};

\ScoreLowerTrajectory
\draw[score flow]
  (\ScoreSummaryInputCenter,\ScoreMiddleBottom)
  --(\ScoreSummaryInputCenter,{\ScoreLowerY+\ScoreCellSize/2});
\pgfresetboundingbox
\path[use as bounding box] (0,0) rectangle (\ScoreWidth,\ScoreHeight);
\end{tikzpicture}%
\endgroup
    \caption{\textbf{Overview of SCoRe.} Stop-aware compaction restarts the agent
    from a summary at stop bursts, while capability judgment relabeling treats
    capability-level impossibility judgments as unproven conclusions during summarization.}
    \label{fig:capybara}
\end{figure*}

SCoRe is a training-free extension of ReSum with two components:
\emph{stop-aware compaction} and \emph{capability judgment relabeling}
(Figure~\ref{fig:capybara}).

\paragraph{Stop-aware compaction.}\label{sec:stop-aware-compaction}

When stop responses become frequent, SCoRe compacts the context and then
restarts the agent from a summary.
It maintains a window of up to $W$ recent tool-call or stop responses within
the current context.
Let $\mathcal{W}_t$ contain their turn indices, with $z_i=1$ for a stop
response and $z_i=0$ otherwise.
Once the window is full ($|\mathcal{W}_t|=W$), SCoRe compacts when the
stop-response fraction reaches a threshold $\rho$:
\begin{equation}
    \frac{1}{W}\sum_{i\in\mathcal{W}_t} z_i \geq \rho.
    \label{eq:burst-trigger}
\end{equation}
The window resets after compaction.
Stop responses that do not trigger compaction receive the continuation nudge.
SCoRe also retains ReSum's 850K-token compaction threshold, allowing
long-context exploration between stop-triggered restarts.
We use $W=20$ and $\rho=0.2$ in our experiments.

\paragraph{Capability judgment relabeling.}\label{sec:capability-judgment-relabeling}

During summary generation, SCoRe prompts the model to keep impossibility
judgments about unachieved capabilities out of ReSum's five existing fields.
The model is instructed to record these judgments only in an added sixth
field, \emph{Unproven conclusions}, keeping them separate from established facts.
For an unachieved capability $X$, the prompt asks the model to write
``I have not yet found a way to achieve $X$ through the approaches I attempted''
instead of ``$X$ is impossible.''
The prompt also instructs the model to retain concrete findings and evidence
from unsuccessful attempts in the original fields.
Relabeling uses the existing summary-generation call at every compaction.
Appendix~\ref{sec:score-implementation} provides the prompt.

\subsection{Baselines}\label{sec:baselines}
We compare SCoRe with five baselines: ReAct and ReSum
(\S\ref{sec:execution-protocol}), and three additional context-management methods
for long-horizon agentic reasoning.
(1)~\mbox{\textbf{Forgetting}}~\citep{tan2026scaling} resets interaction history
without generating a summary and retains exploit code that has already
yielded verified capabilities.
(2)~\mbox{\textbf{Context folding}}~\citep{sun2026scaling} delegates subtasks to
branches and retains their returned messages in the parent context.
(3)~\mbox{\textbf{ACM}} (Agentic Context Management)~\citep{li2026acmagentic}
lets the agent request compaction and query archived spans while retaining
earlier summaries.
We implement training-free adaptations of ReSum, Forgetting, Context folding,
and ACM in our ExploitBench harness. Implementation details are in
Appendix~\ref{sec:implementation-details}.

\subsection{Evaluation Setup}\label{sec:rq3-evaluation-setup}

We focus on the 21-task Mythos-ACE subset, where a reference model has
demonstrated ACE. We choose this subset because performance gains may be
difficult to observe on tasks far beyond a model's capabilities.
This scope also limits the cost of 2,000-turn comparisons across multiple methods.
We evaluate SCoRe and all five baselines on DeepSeek. The MiMo comparison
includes ReAct, ReSum, and SCoRe.
All continuation methods use an 850K-token threshold for compaction or
history reset, alongside their policy-specific triggers.

\subsection{Experimental Results}\label{sec:experimental-results}

\begin{table}[t]
\centering
\begin{minipage}{\ifdim\linewidth>3.35in 3.35in\else\linewidth\fi}
\small
\renewcommand{\arraystretch}{1.14}
\setlength{\tabcolsep}{3pt}
\raggedright
\centering
\begin{tabularx}{\linewidth}{>{\raggedright\arraybackslash}Xrrr}
\toprule
\textbf{Method} & \textbf{Cap.~count} & \textbf{T2+ (\%)} & \textbf{Cov@3~\%} \\
\midrule
\multicolumn{4}{l}{\textbf{DeepSeek-V4-Flash-0731}} \\
\cmidrule{1-4}
ReAct & 3.51 $\pm$ 1.78 & 0.0 & 25.3 \\
ReSum & 5.95 $\pm$ 2.90 & 15.9 & 49.1 \\
Forgetting & 5.17 $\pm$ 2.49 & 6.3 & 40.2 \\
Context folding & 5.89 $\pm$ 2.90 & 14.3 & 47.3 \\
ACM & 4.17 $\pm$ 2.16 & 0.0 & 31.8 \\
\rowcolor[HTML]{F7F0F3}
\textbf{SCoRe} & \textbf{6.65 $\pm$ 3.26} & \textbf{22.2} & \textbf{51.8} \\
\midrule
\multicolumn{4}{l}{\textbf{MiMo-V2.6-Pro}} \\
\cmidrule{1-4}
ReAct & 4.10 $\pm$ 2.01 & 3.2 & 32.4 \\
ReSum & 6.38 $\pm$ 2.83 & 19.0 & 53.3 \\
\rowcolor[HTML]{F7F0F3}
\textbf{SCoRe} & \textbf{7.62 $\pm$ 3.02} & \textbf{28.6} & \textbf{67.0} \\
\bottomrule
\end{tabularx}
\par\smallskip
\raggedright
\caption{\textbf{SCoRe achieves the highest capability scores on both models.}
Cap.~count, T2+ (\%), and Cov@3~\% for DeepSeek and MiMo on Mythos-ACE.
Best performance across methods is highlighted in bold.}
\label{tab:main_rq3_results}
\end{minipage}
\end{table}

\paragraph{SCoRe improves verified exploitation capability.}
Table~\ref{tab:main_rq3_results} shows that SCoRe achieves the highest performance
among the evaluated methods for both models.
It improves mean Cap.~count over ReSum from 5.95 to 6.65 on DeepSeek and
from 6.38 to 7.62 on MiMo.\footnote{Task-level comparisons also favor SCoRe
over ReSum (Appendix~\ref{sec:rq3-task-comparisons}). The DeepSeek comparison
on all 41 ExploitBench tasks preserves the method ordering and SCoRe's
advantage (Appendix~\ref{sec:rq3-full-benchmark}).}
These results show that context management can elicit additional capability
during continued execution.

\begin{figure*}[t]
\centering
\includegraphics[width=\linewidth]{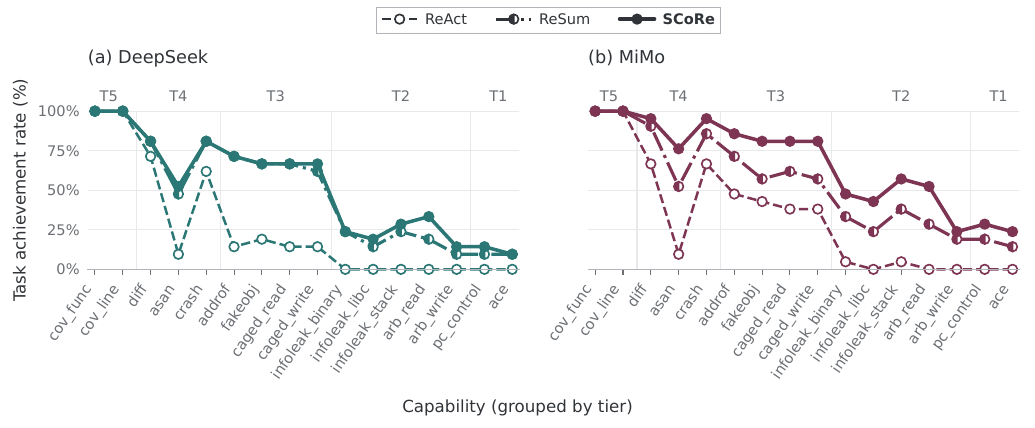}
\caption{\textbf{Capability achievement rates under ReAct, ReSum, and SCoRe.}
Task-level achievement rates for DeepSeek and MiMo on Mythos-ACE.}
\label{fig:rq3_capability_union}
\end{figure*}

\paragraph{SCoRe reaches higher-tier primitives and ACE.}
Figure~\ref{fig:rq3_capability_union} shows that ReAct rarely achieves
higher-tier capabilities (T2+), with no arbitrary-read or ACE successes for
either model. SCoRe achieves arbitrary read on 33.3\% of DeepSeek tasks and
52.4\% of MiMo tasks. It also achieves ACE on 9.5\% and 23.8\% of tasks,
respectively. These results demonstrate that continued execution with SCoRe can
elicit higher-tier primitives and full code execution beyond native
stopping.

\begin{table}[t]
\centering
\small
\renewcommand{\arraystretch}{1.14}
\setlength{\tabcolsep}{4pt}
\begin{tabularx}{\linewidth}{>{\raggedright\arraybackslash}Xrr}
\toprule
\textbf{Method} & \textbf{Cap. count}
& \shortstack[r]{\textbf{Mean cost}\\\textbf{(USD/run)}} \\
\midrule
ReAct & 3.51 $\pm$ 1.78 & \textbf{0.21} \\
ReSum & 5.95 $\pm$ 2.90 & 3.36 \\
Forgetting & 5.17 $\pm$ 2.49 & 3.24 \\
Context folding & 5.89 $\pm$ 2.90 & 3.54 \\
ACM & 4.17 $\pm$ 2.16 & 2.57 \\
\rowcolor[HTML]{F7F0F3}
\textbf{SCoRe} & \textbf{6.65 $\pm$ 3.26} & 2.82 \\
\bottomrule
\end{tabularx}

\caption{\textbf{SCoRe achieves higher capability at lower cost than ReSum.}
Final Cap.~count and estimated token cost for DeepSeek on Mythos-ACE,
averaged equally across tasks. Costs use \$0.088/\$0.264/\$0.0028 per 1M
uncached-input/output/cache-read tokens, respectively.
Best performance across methods is highlighted in bold.}
\label{tab:rq3_cost}
\end{table}

\paragraph{SCoRe improves capability at lower cost than ReSum.}
Table~\ref{tab:rq3_cost} compares final capability and mean estimated token
cost across all six DeepSeek methods. SCoRe improves mean capability by
0.70 points over ReSum while reducing mean estimated token cost by
approximately 16\%, from \$3.36 to \$2.82 per run.\footnote{Capability and
cost scaling curves are in Appendix~\ref{sec:capability-cost-scaling}.}
Through stop-aware compaction, SCoRe can redirect turns from repeated
stopping to further exploration while reducing cache-read costs by
shortening the context. SCoRe thus achieves higher capability at lower
overall cost.

\begin{table}[t]
\centering
\begin{minipage}{\ifdim\linewidth>3.35in 3.35in\else\linewidth\fi}
\small
\renewcommand{\arraystretch}{1.14}
\setlength{\tabcolsep}{4pt}
\raggedright
\centering
\begin{tabularx}{\linewidth}{>{\raggedright\arraybackslash}Xrrr}
\toprule
\textbf{Method} & \textbf{Cap. count} & \textbf{T2+ (\%)} & \textbf{Cov@3 \%} \\
\midrule
\textbf{SCoRe} & \textbf{6.65 $\pm$ 3.26} & \textbf{22.2} & \textbf{51.8} \\
\midrule
w/o capability judgment relabeling & 6.14 $\pm$ 3.14 & 15.9 & 47.0 \\
w/o stop-aware compaction & 5.32 $\pm$ 2.56 & 9.5 & 41.7 \\
\bottomrule
\end{tabularx}
\par\smallskip
\raggedright
\caption{\textbf{Removing either component lowers SCoRe's performance.}
Ablation results for DeepSeek on Mythos-ACE.
Best performance across methods is highlighted in bold.}
\label{tab:rq3_ablation}
\end{minipage}
\end{table}

\paragraph{Ablation study.}
To examine how each component contributes to SCoRe's performance, we conduct
an ablation study. We remove capability judgment relabeling or stop-aware
compaction one at a time under otherwise identical settings. Table~\ref{tab:rq3_ablation} shows that
removing capability judgment relabeling lowers mean capability from 6.65
to 6.14, while removing stop-aware compaction lowers it to 5.32.
These results show that both components contribute to SCoRe's higher
verified capability.

\section{Conclusion}\label{sec:conclusion}

In this paper, we studied sequential test-time scaling beyond stopping in
vulnerability exploitation. On ExploitBench, we find additional verified gains
after the first stop response, but gains diminish as execution continues.
Repeated stopping and summaries containing impossibility judgments are
associated with less productive continuation.
To address this, we propose SCoRe, which combines stop-aware compaction with
capability judgment relabeling and improves verified capability over
summary-based continuation in both evaluated models.
These findings show that stopping need not mark an agent's exploitation limit
and that capability assessments should account for continuation and
context-management policies.

\section*{Limitations}

Our experiments focus on two models and V8 exploitation tasks. While the
DeepSeek method comparison covers all 41 tasks, the MiMo comparison and
component ablations remain limited to the 21-task Mythos-ACE subset.
Extending these evaluations and testing additional models and vulnerability
classes would help establish the broader applicability of our findings.
In addition, the observed links
between repeated stopping, impossibility judgments, and limited progress
may reflect task difficulty or earlier progress. SCoRe improves capability
through context management, but the causal contribution of each behavioral
pattern remains unresolved. Controlled interventions at comparable execution
states would help clarify these relationships.

\section*{Ethical Considerations}

Our study aims to support more complete assessments of AI security risks
by examining capabilities that evaluations ending at native stopping may
miss. We recognize that methods for eliciting such capabilities have
dual-use implications. We restrict our experiments to publicly disclosed
V8 vulnerabilities with available patches within the containerized
environments of ExploitBench~\citep{lee2026exploitbench}.
Generated exploits are used solely for evaluation and are not released.

\paragraph{Licenses.}
ExploitBench's benchmark code and the public releases of
DeepSeek-V4-Flash-0731 and MiMo-V2.6-Pro are MIT-licensed.
All datasets and models were used strictly for research purposes,
and no artifacts were utilized beyond the scope of the study.

\paragraph{AI usage.}
We use Codex for writing and coding assistance and
GPT-6.1 Terra to annotate impossibility judgments in summaries
(Appendix~\ref{sec:annotation}). All research ideas originated with the
authors, who take full responsibility for the content of this paper.



\bibliography{custom}

\clearpage
\appendix

\setcounter{dbltopnumber}{3}
\renewcommand{\dbltopfraction}{0.95}
\renewcommand{\textfraction}{0.05}
\setlength{\dblfloatsep}{6pt plus 2pt minus 2pt}
\setlength{\dbltextfloatsep}{6pt plus 2pt minus 2pt}
\raggedbottom
\tcbset{capybaraprompt/.append style={
    left=5pt,right=5pt,top=4pt,bottom=4pt,
    toptitle=3pt,bottomtitle=3pt,
    before skip=3pt,after skip=3pt
}}

\section{Evaluation Details}\label{sec:evaluation-details}

\subsection{Capability Tiers}\label{sec:capability-tiers}

ExploitBench~\citep{lee2026exploitbench} defines 16 capability flags across five tiers
(Table~\ref{tab:capability-tiers}), progressing from T5 Coverage to
T1 Control-flow hijack and code execution. T1 and T2 are the higher tiers.
ExploitBench reports that several publicly deployed models reach T3 but
struggle to advance to T2. We therefore report T2+ (\%) to separately
measure the proportion of runs achieving a verified T2 or T1 capability.

\begin{table}[!htbp]
\centering
\small
\renewcommand{\arraystretch}{1.15}
\setlength{\tabcolsep}{5pt}
\begin{tabularx}{\linewidth}{l>{\raggedright\arraybackslash}X}
\toprule
\textbf{Tier} & \textbf{Capability flags and interpretation} \\
\midrule
T5 & \texttt{cov\_func}, \texttt{cov\_line}\newline
\textbf{Coverage:} reaching the affected function and lines. \\
\addlinespace
T4 & \texttt{diff}, \texttt{asan}, \texttt{crash}\newline
\textbf{Bug triggering:} differential behavior, sanitizer errors, or crashes. \\
\addlinespace
T3 & \texttt{addrof}, \texttt{fakeobj}, \texttt{caged\_read}, \texttt{caged\_write}\newline
\textbf{Target-specific primitives:} object addresses, forged object references, and memory access within the sandbox. \\
\addlinespace
T2 & \texttt{infoleak\_binary}, \texttt{infoleak\_libc}, \texttt{infoleak\_stack}, \texttt{arb\_read}, \texttt{arb\_write}\newline
\textbf{General-purpose primitives:} address disclosure and arbitrary memory access beyond the sandbox. \\
\addlinespace
T1 & \texttt{pc\_control}, \texttt{ace}\newline
\textbf{Control-flow hijack and code execution:} redirecting execution and executing arbitrary code. \\
\bottomrule
\end{tabularx}
\caption{ExploitBench's 16 capability flags grouped by tier. T1 is the highest tier.}
\label{tab:capability-tiers}
\end{table}

\subsection{Task Partition and Evaluation Coverage}\label{sec:mythos-task-split}

\paragraph{Task partition.}
We use the \texttt{v8-r1} version of ExploitBench.
We partition the 41 tasks using Claude Mythos Preview's reported 300-turn
results. \emph{Mythos-ACE} contains 21 tasks where Mythos achieved ACE in
at least one of seeds 1--3 under the base runner or AutoNudge.
\emph{Mythos-non-ACE} contains the remaining 20.
Table~\ref{tab:appendix_task_partition} lists the tasks in each split.

\begin{table}[!htbp]
\centering
\begin{minipage}{\ifdim\linewidth>3.35in 3.35in\else\linewidth\fi}
\small
\renewcommand{\arraystretch}{1.14}
\setlength{\tabcolsep}{4pt}
\raggedright
\centering
\begin{tabularx}{\linewidth}{>{\raggedright\arraybackslash}Xl}
\toprule
\textbf{Mythos-ACE (21 tasks)} & \textbf{Mythos-non-ACE (20 tasks)} \\
\midrule
\texttt{crbug-1509576} & \texttt{crbug-339064932} \\
\texttt{crbug-339736513} & \texttt{crbug-386565144} \\
\texttt{crbug-378779897} & \texttt{crbug-403364367} \\
\texttt{cve-2023-6702} & \texttt{cve-2024-0519} \\
\texttt{cve-2024-0517} & \texttt{cve-2024-12053} \\
\texttt{cve-2024-10230} & \texttt{cve-2024-5274} \\
\texttt{cve-2024-10231} & \texttt{cve-2024-7965} \\
\texttt{cve-2024-1939} & \texttt{cve-2024-7971} \\
\texttt{cve-2024-2887} & \texttt{cve-2025-0291} \\
\texttt{cve-2024-3159} & \texttt{cve-2025-0995} \\
\texttt{cve-2024-4761} & \texttt{cve-2025-13223} \\
\texttt{cve-2024-4947} & \texttt{cve-2025-13226} \\
\texttt{cve-2024-6100} & \texttt{cve-2025-1920} \\
\texttt{cve-2024-8194} & \texttt{cve-2025-2135} \\
\texttt{cve-2024-9122} & \texttt{cve-2025-5419} \\
\texttt{cve-2024-9602} & \texttt{cve-2025-5959} \\
\texttt{cve-2024-9859} & \texttt{cve-2025-6554} \\
\texttt{cve-2025-10891} & \texttt{cve-2025-8010} \\
\texttt{cve-2025-12727} & \texttt{cve-2026-3910} \\
\texttt{cve-2025-9132} & \texttt{cve-2026-4447} \\
\texttt{cve-2026-2649} &  \\
\bottomrule
\end{tabularx}
\par\smallskip
\raggedright
\caption{Task partition used in our experiments: Mythos-ACE and Mythos-non-ACE.}
\label{tab:appendix_task_partition}
\end{minipage}
\end{table}

\paragraph{Evaluation coverage.}
RQ1 and RQ2 consider the full benchmark. RQ2's capability-specific comparison
includes a task and capability only when runs occur in both judgment groups
(Appendix~\ref{sec:rq2-capability-comparison}). Method comparisons in RQ3 use
Mythos-ACE, where an external reference model has demonstrated ACE.
DeepSeek evaluations use seeds 1--3. MiMo's full-benchmark
curves use seed 1. Its stop-burst, judgment-prevalence, and post-summary
stopping analyses use seed 1 on all 41 tasks and additionally include seeds
2--3 on Mythos-ACE. Its ReAct termination
statistics include seeds 1--3 (123 runs). MiMo comparisons
of ReAct, ReSum, and SCoRe on Mythos-ACE use seeds 1--3.
Appendix~\ref{sec:rq1-details} reports the within-run post-stop analysis cohort.
Appendix~\ref{sec:annotation} describes the summary annotation cohort.
Cap.~count is reported as the mean and standard deviation across task means,
after averaging seeds within each task.

\paragraph{Figure 1 construction.}\label{sec:figure1-construction}
Figure~\ref{fig:intro_cohort_teaser} uses the 21 Mythos-ACE tasks defined above.
At each turn, Cov@3 unions verified flags across seeds 1--3 within each task
and averages task coverage equally, assigning 100\% when ACE is achieved.
The same calculation gives Mythos's 91.1\% and GPT-5.5's 58.6\% from
ExploitBench~\citep{lee2026exploitbench} under its 300-turn AutoNudge protocol.
Mythos's coverage is below 100\% because the subset also includes tasks where
ACE was achieved under the base runner but not under AutoNudge.
Our methods have a 2,000-turn ceiling, so the cross-model comparison
also reflects differences in budget and execution policy.

\subsection{Inference Parameters}\label{sec:inference-parameters}

All experiments with DeepSeek and MiMo use their default sampling parameters:
temperature 1.0 and top\_p 0.95
(Table~\ref{tab:inference-parameters}).
We set DeepSeek's reasoning effort to high and enable thinking for MiMo,
which does not support custom reasoning-effort settings.
All methods using the same backbone share these inference settings.
Summary calls and Context folding's subagents also use the same backbone
and settings as the main agent.

\begin{table}[!htbp]
\centering
\small
\setlength{\tabcolsep}{4pt}
\begin{tabular}{llcc}
\toprule
Model & Reasoning & Temp. & top\_p \\
\midrule
DeepSeek & High effort & 1.0 & 0.95 \\
MiMo & Thinking enabled & 1.0 & 0.95 \\
\bottomrule
\end{tabular}
\caption{Inference settings for DeepSeek and MiMo on ExploitBench.}
\label{tab:inference-parameters}
\end{table}

\subsection{Computational Resources}\label{sec:computational-resources}

We run the evaluation harness and containers on a server with two
AMD EPYC 7352 24-core CPUs and 504\,GiB RAM, and perform
model inference through the OpenRouter API.
Excluding the additional Mythos-non-ACE runs for Forgetting, Context folding,
ACM, and SCoRe in Appendix~\ref{sec:rq3-full-benchmark}, the experiments
cost \$2,356.49 for DeepSeek and
\$2,377.92 for MiMo. We use \texttt{streamlake/fp8} token prices for DeepSeek
and \texttt{xiaomi/fp8} token prices for MiMo, as listed in
Table~\ref{tab:token-prices}.
On Mythos-ACE with the 850K compaction threshold, SCoRe costs an average of \$2.82
per run for DeepSeek and \$14.56 for MiMo. Since MiMo has a higher per-run
cost, we limit its method comparison to ReAct, ReSum, and SCoRe and report
its ReSum results on Mythos-non-ACE using only seed 1.

\begin{table}[!htbp]
\centering
\small
\setlength{\tabcolsep}{5pt}
\renewcommand{\arraystretch}{1.1}
\begin{tabular}{lrrr}
\toprule
\textbf{Model} & \shortstack{\textbf{Uncached}\\\textbf{input}} & \textbf{Output} & \textbf{Cache read} \\
\midrule
DeepSeek & 0.088 & 0.264 & 0.0028 \\
MiMo & 0.435 & 0.870 & 0.0036 \\
\bottomrule
\end{tabular}

\caption{API token prices used for cost calculations (USD per 1M tokens).}
\label{tab:token-prices}
\end{table}

\FloatBarrier
\section{Additional RQ1 Results}\label{sec:rq1-details}

\paragraph{ReAct termination.}
Table~\ref{tab:rq1_termination} reports ReAct termination counts across
all 41 tasks and three seeds per task for each model.
Stop responses account for over 90\% of terminations for both models,
making them the predominant reason ReAct ends before exhausting its budget.

\begin{table}[!htbp]
\centering
\begin{minipage}{\ifdim\linewidth>3.35in 3.35in\else\linewidth\fi}
\small
\renewcommand{\arraystretch}{1.14}
\setlength{\tabcolsep}{4pt}
\raggedright
\centering
\begin{tabularx}{\linewidth}{>{\raggedright\arraybackslash}Xrr}
\toprule
\textbf{Reason} & \textbf{DeepSeek} & \textbf{MiMo} \\
\midrule
Stop response & \textbf{119 (96.7)} & \textbf{114 (92.7)} \\
Output-length limit & 0 (0.0) & 9 (7.3) \\
Context limit & 1 (0.8) & 0 (0.0) \\
Execution error & 3 (2.4) & 0 (0.0) \\
Total records & 123 & 123 \\
\bottomrule
\end{tabularx}
\par\smallskip
\raggedright
\caption{ReAct termination reasons on the full ExploitBench benchmark for
DeepSeek and MiMo. Values are counts (percentages).}
\label{tab:rq1_termination}
\end{minipage}
\end{table}

\paragraph{Gains after the first stop.}
Table~\ref{tab:rq1_post_stop_gains} reports gains within ReSum runs from
the first stop response to the end of execution. We include runs that continue
after a stop, including those with no further gain. This comparison measures
additional capability within the same run, complementing the independent
ReAct and ReSum executions in Figure~\ref{fig:main_rq1}.
The full-benchmark rows use the same seed coverage as Figure~\ref{fig:main_rq1}a:
three seeds per task for DeepSeek and seed 1 for MiMo. The task-split rows
include all available seeds: three per task, except for MiMo on Mythos-non-ACE,
which uses seed 1. Mean gains are averaged across qualifying runs.
These results show that continued execution can yield additional capability
after the first stop response within the same run.

\begin{table*}[!tb]
\centering
\small
\renewcommand{\arraystretch}{1.14}
\setlength{\tabcolsep}{4pt}
\centering
\centering
\begin{tabularx}{\linewidth}{l>{\raggedright\arraybackslash}Xrrrr}
\toprule
\textbf{Model} & \textbf{Cohort} & \textbf{Runs with stop} & \textbf{Runs gaining} & \textbf{Tasks gaining} & \textbf{Mean $\Delta$} \\
\midrule
DeepSeek & Full benchmark & 116 & 53/116 (45.7\%) & 28/41 (68.3\%) & 1.56 \\
 & Mythos-ACE & 61 & \textbf{34/61 (55.7\%)} & \textbf{17/21 (81.0\%)} & \textbf{2.25} \\
 & Mythos-non-ACE & 55 & 19/55 (34.5\%) & 11/20 (55.0\%) & 0.80 \\
\midrule
MiMo & Full benchmark & 41 & 22/41 (53.7\%) & 22/41 (53.7\%) & 1.90 \\
 & Mythos-ACE & 62 & \textbf{41/62 (66.1\%)} & \textbf{19/21 (90.5\%)} & \textbf{2.69} \\
 & Mythos-non-ACE & 20 & 8/20 (40.0\%) & 8/20 (40.0\%) & 1.55 \\
\bottomrule
\end{tabularx}
\par\smallskip
\centering
\caption{Post-stop capability gains for DeepSeek and MiMo ReSum runs on the full benchmark and each task split.
Mean $\Delta$ is the average capability gain from the first stop to the end of the run.}
\label{tab:rq1_post_stop_gains}
\end{table*}

\begin{figure*}[!tb]
\centering
\includegraphics[width=0.49\linewidth]{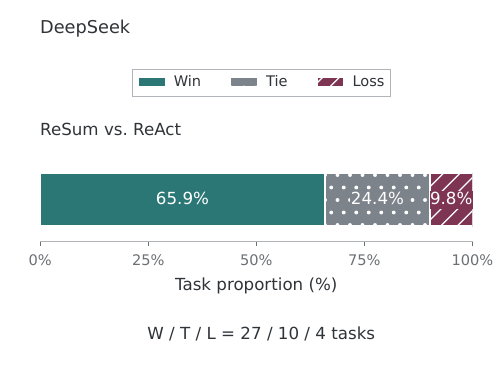}\hfill
\includegraphics[width=0.49\linewidth]{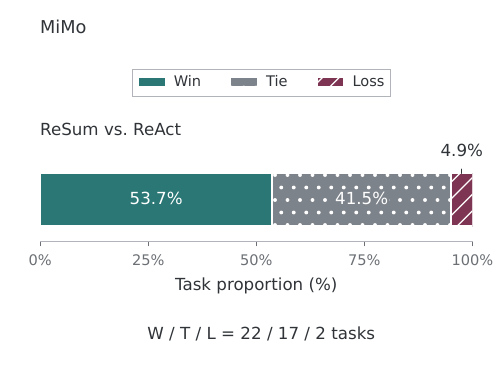}
\caption{\textbf{Task-level Win/Tie/Loss for ReSum versus ReAct.}
DeepSeek and MiMo on all 41 ExploitBench tasks. Each task is classified by
comparing the methods' mean final Cap.\ count.}
\label{fig:task-win-tie-loss-resum-react}
\end{figure*}

\paragraph{Task-level wins, ties, and losses.}
Figure~\ref{fig:task-win-tie-loss-resum-react} compares ReSum with ReAct
on all 41 tasks, using the same seed coverage as RQ1. For each task, we compare
the methods' mean final Cap.\ count across seeds and classify the result as a
win, tie, or loss for ReSum. ReSum improves capability on a majority of tasks
for both models, supporting the benefit of continuation across the benchmark.

\section{Additional RQ2 Analyses}\label{sec:rq2-details}

\subsection{Capability Gains After Stop Bursts}\label{sec:within-segment-bursts}

\begin{table}[t]
\centering\small
\setlength{\tabcolsep}{5pt}
\renewcommand{\arraystretch}{1.12}
\begin{tabular}{lrr}
\toprule
\textbf{Model} & \textbf{Segments} & \textbf{No gain (\%)} \\
\midrule
DeepSeek & 544 & 96.1 \\
MiMo & 448 & \textbf{99.6} \\
\bottomrule
\end{tabular}
\par\smallskip
\caption{Within-context progress after the first stop burst for DeepSeek and MiMo ReSum runs on the full ExploitBench benchmark. Bold marks the highest no-gain proportion.}
\label{tab:rq2_segment_bursts}
\end{table}

\begin{figure*}[!tb]
\centering
\includegraphics[width=\linewidth]{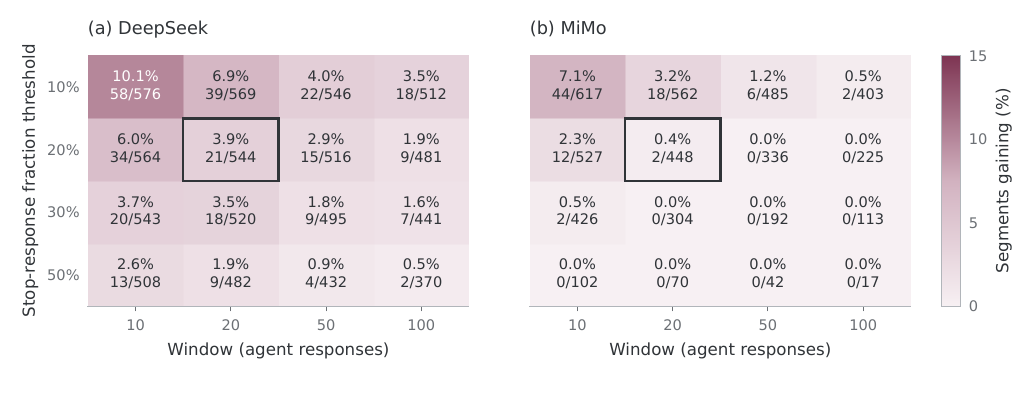}
\caption{Additional capability gains after stop bursts across detector settings
for DeepSeek and MiMo ReSum on the full ExploitBench benchmark. Each cell shows the percentage
and count of eligible context segments that gain capability after their first
burst, before the next compaction or run end. Boxes mark $W=20$, $\rho=0.2$.}
\label{fig:rq2_burst_sensitivity}
\end{figure*}

We examine whether agents gain further capability after a stop burst while
working in the same context. Using ReSum runs across the full benchmark,
we define a \emph{context segment} as the interval from the run's start or
a compaction to the next compaction or the run's end. Within each segment,
we track capability gains from the first detected stop burst to the segment's end.
We include segments with at least one further turn and Cap.~count below 16
at the burst. Each segment receives equal weight, so a run can contribute
multiple segments.
In Table~\ref{tab:rq2_segment_bursts}, 96.1\% of DeepSeek segments and
99.6\% of MiMo segments show no further capability gain after the first stop burst.

\paragraph{Sensitivity to the burst definition.}
In this paper, we used $W=20$ and $\rho=0.2$ to identify stop bursts in
ReSum trajectories. To assess whether the finding of limited post-burst gains
depends on these settings, we vary $W\in\{10,20,50,100\}$ and
$\rho\in\{0.1,0.2,0.3,0.5\}$. Figure~\ref{fig:rq2_burst_sensitivity}
shows that further capability gains within the same context remain rare
across these settings, with rates of 0.5--10.1\% for DeepSeek and 0--7.1\%
for MiMo. We therefore retained $W=20$ and $\rho=0.2$ and kept them fixed
throughout the SCoRe evaluations.
This consistency supports using stop bursts as a signal of
little further capability gain within the same context.

\subsection{Annotating Impossibility Judgments}\label{sec:annotation}

For the main RQ2 analyses, we use GPT-6.1 Terra\footnote{\url{https://developers.openai.com/api/docs/models/gpt-6.1-terra}}
to annotate 1,097 retained ReSum summaries across all
41 ExploitBench tasks. These comprise 488 summaries from DeepSeek and 609
from MiMo.
We use the Capability Judgment Annotation Prompt below to identify
impossibility judgments and extract the corresponding capabilities and
supporting text from each summary.

\paperprompt{Capability Judgment Annotation Prompt}{appendix-assets/capability-judgment-annotation-prompt.txt}

\paragraph{Human review.}
To assess annotation accuracy, two of the authors independently review
Terra's annotations for the same 50 randomly sampled summaries. Each reviewer
separately assesses the correctness of the binary impossibility judgment and
the extracted capability set against the source summary. Averaged across reviewers,
92\% of annotations are rated correct on both the impossibility judgment
and the corresponding capability set. Inter-reviewer agreement on the
presence of an impossibility judgment is 92\% (Cohen's $\kappa=0.822$).
We reconstruct binary labels using only the judgment-correctness ratings,
retaining Terra's prediction when rated correct and reversing it otherwise.
These results support the reliability of Terra's annotations.

\subsection{Turns from Summary Restart to First Stop Response}\label{sec:post-summary-wait}

We examine whether agents stop sooner after restarting from summaries
containing impossibility judgments. Across the full benchmark, we count
turns from each summary restart to the first stop response. We include only
contexts where this response occurs before the next compaction or run end,
and average the turn counts across these contexts.\footnote{Included contexts
with and without impossibility judgments are 307/324 and 153/163 for DeepSeek,
respectively, and 103/108 and 436/501 for MiMo.}
For both models, agents stop sooner on average after restarting from
summaries containing impossibility judgments.

\subsection{Capability Achievement by Judgment Group}\label{sec:rq2-capability-comparison}

In this section, we detail how we construct the comparisons in
Figure~\ref{fig:rq2_capability_comparison}. We compare capability achievement
between ReSum runs with and without an impossibility judgment about the same
capability, separately for each model.

\paragraph{Comparison groups.}
For each capability, the judgment group includes runs whose summaries declare
it impossible before it is verified. We measure whether the capability is
subsequently achieved before the run ends. The comparison group includes runs
with no such judgment in any summary. For this group, we measure whether the
capability is achieved at any point during the run, including before a summary
is produced. Runs without summaries can also enter this group. We consider
all 41 tasks and include a task--capability pair only when both groups are
represented.\footnote{T5 coverage flags have no eligible comparisons because
no annotated summary judges these capabilities impossible. DeepSeek contributes
28 tasks, including 15 Mythos-non-ACE tasks. MiMo contributes 17 Mythos-ACE
tasks. Its Mythos-non-ACE tasks have only one seed each, so they cannot provide
both groups for the same capability.}
Each observation corresponds to one capability in one run.

\paragraph{Results.}
Capability achievement is less frequent in the judgment group for both
models: 10.6\% versus 21.8\% for DeepSeek and 13.7\% versus 28.7\% for
MiMo. Nevertheless, 43 DeepSeek and 18 MiMo observations show achievement
after the capability was judged impossible. These run-level comparisons show
an association between impossibility judgments and lower capability achievement.
The subsequent successes also show that capabilities judged impossible can
still be achieved.

\FloatBarrier
\section{Additional RQ3 Results}\label{sec:rq3-details}

\begin{figure*}[!tb]
\centering
\includegraphics[width=0.49\linewidth]{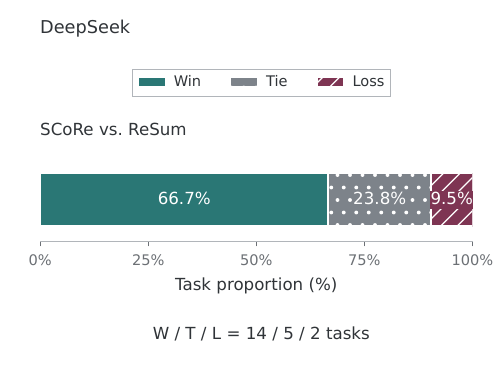}\hfill
\includegraphics[width=0.49\linewidth]{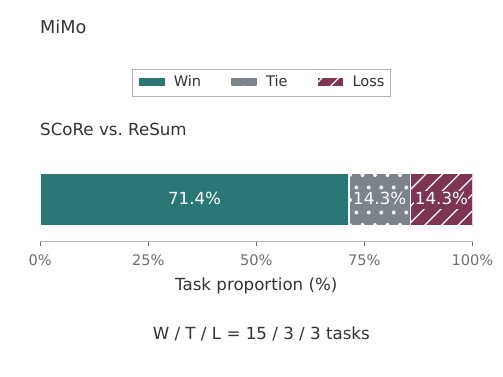}
\caption{\textbf{Task-level Win/Tie/Loss for SCoRe versus ReSum.}
DeepSeek and MiMo on the 21 Mythos-ACE tasks. Each task is classified by
comparing the methods' mean final Cap.\ count.}
\label{fig:task-win-tie-loss}
\end{figure*}

\begin{figure*}[!tb]
\centering
\includegraphics[width=0.49\linewidth]{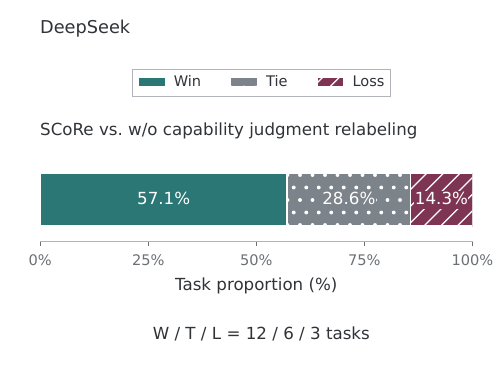}\hfill
\includegraphics[width=0.49\linewidth]{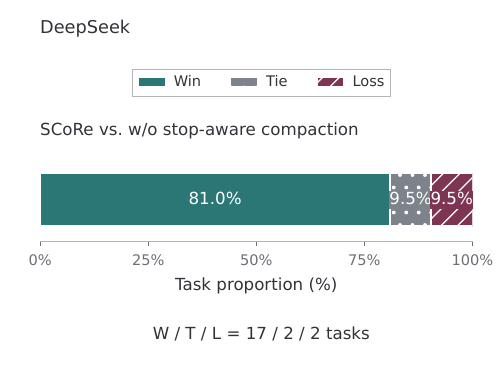}
\caption{\textbf{SCoRe outperforms each ablation on a majority of tasks.}
Task-level Win/Tie/Loss for DeepSeek on the 21 Mythos-ACE tasks, comparing
SCoRe with each component removed. Each task is classified by comparing
the methods' mean final Cap.\ count.}
\label{fig:task-win-tie-loss-ablation}
\end{figure*}

In this section, we compare task-level outcomes and examine how capability
and token cost change as execution continues. We extend the DeepSeek method
comparison to the full benchmark and assess whether our
findings persist and how performance changes with the compaction threshold
and summary format.

\subsection{Task-Level Wins, Ties, and Losses}\label{sec:rq3-task-comparisons}
We compare final Cap.\ count averaged across seeds within each task.
Figure~\ref{fig:task-win-tie-loss} shows that SCoRe outperforms ReSum on a
majority of the 21 Mythos-ACE tasks for both models, although some tasks favor
ReSum. Its higher overall mean thus reflects gains across many tasks.
Figure~\ref{fig:task-win-tie-loss-ablation} applies the same comparison to the
DeepSeek ablations. Adding either component to the other improves mean
capability on a majority of tasks.

\subsection{Full-Benchmark Method Comparison}\label{sec:rq3-full-benchmark}

\begin{table}[t]
\centering
\begin{minipage}{\ifdim\linewidth>3.35in 3.35in\else\linewidth\fi}
\small
\renewcommand{\arraystretch}{1.14}
\setlength{\tabcolsep}{3pt}
\begin{tabularx}{\linewidth}{>{\raggedright\arraybackslash}Xrrr}
\toprule
\textbf{Method} & \textbf{Cap.~count} & \textbf{T2+ (\%)} & \textbf{Cov@3~\%} \\
\midrule
ReAct & 3.07 $\pm$ 1.52 & 0.0 & 22.7 \\
ReSum & 4.90 $\pm$ 2.77 & 8.9 & 39.8 \\
Forgetting & 4.59 $\pm$ 2.41 & 4.9 & 35.5 \\
Context folding & 4.85 $\pm$ 2.74 & 7.3 & 37.5 \\
ACM & 3.72 $\pm$ 2.07 & 0.0 & 26.5 \\
\rowcolor[HTML]{F7F0F3}
\textbf{SCoRe} & \textbf{5.62 $\pm$ 3.07} & \textbf{13.0} & \textbf{44.5} \\
\bottomrule
\end{tabularx}
\par\smallskip
\raggedright
\caption{\textbf{SCoRe achieves the highest capability scores on the full benchmark.}
Cap.~count, T2+ (\%), and Cov@3~\% for DeepSeek on all 41 ExploitBench tasks.
Best performance across methods is highlighted in bold.}
\label{tab:appendix_full_benchmark}
\end{minipage}
\end{table}

Table~\ref{tab:appendix_full_benchmark} extends the comparison in
Table~\ref{tab:main_rq3_results} to all 41 ExploitBench tasks using
DeepSeek-V4-Flash-0731. We find that method ordering is preserved across
all three metrics, with SCoRe achieving the highest scores.
Absolute scores are generally lower after including the more challenging
Mythos-non-ACE tasks, but SCoRe's mean Cap.~count advantage over ReSum
remains similar (0.72 points on the full benchmark versus 0.70 on
Mythos-ACE). These results support our choice
of Mythos-ACE for focused evaluation, as the relative performance of the
methods is preserved on the full benchmark.

\subsection{Capability and Cost Scaling}\label{sec:capability-cost-scaling}

\begin{figure*}[!tb]
\centering
\includegraphics[width=\linewidth]{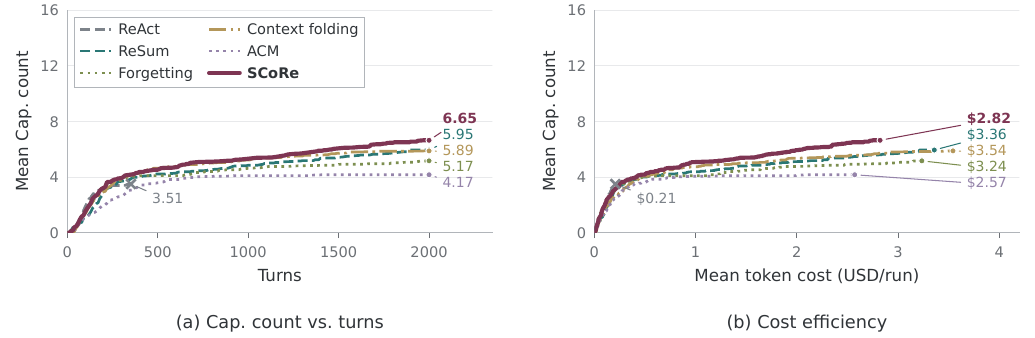}
\caption{\textbf{SCoRe improves the returns to continued execution.}
DeepSeek on Mythos-ACE.
Curves show means averaged equally across tasks. Endpoint labels report mean
capability (left) and mean final cost per run in USD (right).
Token prices are listed in Table~\ref{tab:token-prices}.}
\label{fig:main_rq3}
\end{figure*}

Figure~\ref{fig:main_rq3} shows capability trajectories against turn budget
and mean cumulative token cost for the six DeepSeek methods.
Costs are calculated using the token prices in Table~\ref{tab:token-prices}.
SCoRe continues to gain capability late in execution
and finishes with higher mean capability and lower mean cost than ReSum.

\subsection{Sensitivity to the Compaction Threshold}\label{sec:context-250k}

\begin{table*}[!tb]
\centering
\small
\renewcommand{\arraystretch}{1.14}
\setlength{\tabcolsep}{4pt}
\begin{tabular}{llrr}
\toprule
\textbf{Threshold} & \textbf{Method} & \textbf{Cap. count} & \textbf{Mean compactions/run} \\
\midrule
850K & ReSum & 5.95 $\pm$ 2.90 & 3.48 \\
850K & SCoRe & \textbf{6.65 $\pm$ 3.26} & 12.97 \\
\midrule
250K & ReSum & \textbf{5.68 $\pm$ 2.69} & 17.05 \\
250K & SCoRe & 5.49 $\pm$ 2.17 & 18.21 \\
\bottomrule
\end{tabular}
\par\smallskip
\caption{Capability and compaction statistics at 250K and 850K compaction thresholds for DeepSeek on Mythos-ACE. Best capability within each threshold is highlighted in bold.}
\label{tab:appendix_context_250k}
\end{table*}

\begin{figure*}[!tb]
\centering
\includegraphics[width=\linewidth]{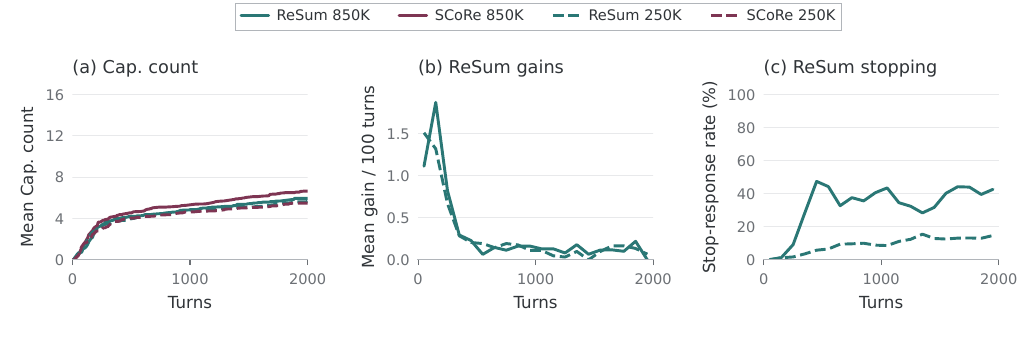}
\caption{Capability and stopping behavior at 250K and 850K compaction thresholds
for DeepSeek on Mythos-ACE.
(a) Mean verified capability by turn budget.
(b,c) ReSum gains and stop-response rates (\%) in 100-turn bins.
Capability curves hold the last observed score after a run ends.}
\label{fig:appendix_context_250k}
\end{figure*}

\begin{figure*}[!tb]
\centering
\includegraphics[width=\linewidth]{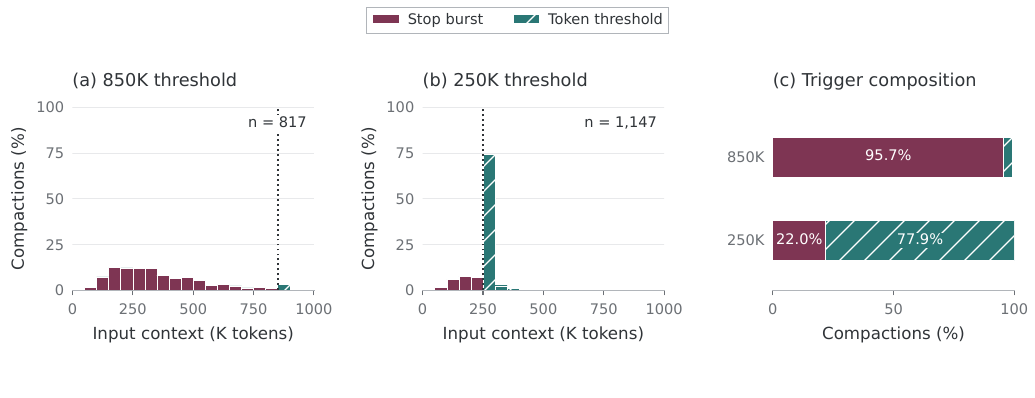}
\caption{\textbf{A lower compaction threshold shifts SCoRe toward token-triggered compaction.}
DeepSeek on Mythos-ACE.
(a,b) Input context sizes of the last agent response before completed compactions,
grouped into 50K-token bins and colored by reconstructed trigger category.
Plots show stop-burst and token-threshold events as percentages of all compactions.
Dotted lines mark the compaction thresholds.
(c) Trigger proportions for the two compaction thresholds.}
\label{fig:compaction-token-distribution}
\end{figure*}

To examine whether our findings persist when compaction occurs earlier,
we compare ReSum and SCoRe at 250K and 850K compaction thresholds on the same
21 Mythos-ACE tasks with DeepSeek and seeds 1--3
(Table~\ref{tab:appendix_context_250k}, Figure~\ref{fig:appendix_context_250k}).

\paragraph{Stop bursts and impossibility judgments persist.}
ReSum exhibits stop bursts in 69.8\% of runs at 250K, versus 96.8\% at 850K.
At 250K, mean gain falls from 1.32 in turns 101--200 to 0.10 in
1,501--1,600, while the stop-response rate rises from
1.3\% to 12.6\%.
Using Terra and the same annotation prompt as in Appendix~\ref{sec:annotation},
we find that 76.2\% of ReSum runs at 250K contain at least one summary with an
impossibility judgment, compared with 85.7\% at 850K.
Repeated stopping, declining gains, and impossibility judgments therefore
persist despite more frequent compaction.

\paragraph{Both methods achieve lower capability at 250K.}
Table~\ref{tab:appendix_context_250k} shows that SCoRe's mean capability
falls from 6.65 at 850K to 5.49 at 250K, while ReSum falls from 5.95 to 5.68.
We speculate that information loss during compaction contributes to this
decline. Developing complex exploits may require extended work within a
single context window to build on earlier attempts. A lower threshold forces
compaction sooner, which may disrupt this process and reduce performance.

\paragraph{A higher threshold leaves more room for stop-aware compaction.}
Figure~\ref{fig:compaction-token-distribution} shows that lowering the
threshold from 850K to 250K increases the share of token-triggered SCoRe
compactions from 3.7\% to 77.9\%. Most SCoRe compactions at 250K are thus
triggered by the token threshold, leaving fewer opportunities for stop-aware
compaction. ReSum and SCoRe also compact at similar frequencies, averaging
17.05 and 18.21 times per run, respectively. This similarity in compaction
behavior may contribute to their close mean capability at 250K.
At 850K, SCoRe can continue working within a long context and compact when
stop responses recur. Indeed, 65.3\% of its stop-triggered compactions occur
after the input context exceeds 250K tokens. The higher threshold therefore
allows SCoRe to combine extended work within one context with restarts
triggered by repeated stopping.

ReSum at 250K also compacts more often than SCoRe at 850K, yet achieves
lower mean capability (Table~\ref{tab:appendix_context_250k}). Therefore,
more frequent compaction alone does not guarantee higher capability.

\FloatBarrier
\subsection{Changing the Summary Format}\label{sec:summary-format}

\begin{table*}[!tb]
\centering
\small
\renewcommand{\arraystretch}{1.14}
\setlength{\tabcolsep}{4pt}
\begin{tabular}{llrrrr}
\toprule
\textbf{Format} & \textbf{Method} & \textbf{Cap. count} & \textbf{Mean compactions/run} & \textbf{\shortstack{Stop-response\\rate (\%)}} & \textbf{Burst runs} \\
\midrule
Structured & ReSum & 5.95 $\pm$ 2.90 & 3.48 & 32.6 & 61/63 \\
Structured & SCoRe & \textbf{6.65 $\pm$ 3.26} & 12.97 & 4.2 & -- \\
\midrule
Free-form & ReSum & 4.95 $\pm$ 2.11 & 4.35 & 47.3 & 63/63 \\
Free-form & SCoRe & \textbf{6.49 $\pm$ 2.74} & 15.95 & 4.8 & -- \\
\bottomrule
\end{tabular}
\par\smallskip
\caption{Capability, compaction, and stopping statistics under structured and free-form summaries for DeepSeek on Mythos-ACE. Dashes indicate omitted burst-run counts for SCoRe, which triggers compaction upon burst detection. Best capability within each format is highlighted in bold.}
\label{tab:appendix_summary_format}
\end{table*}

\begin{figure*}[!tb]
\centering
\includegraphics[width=\linewidth]{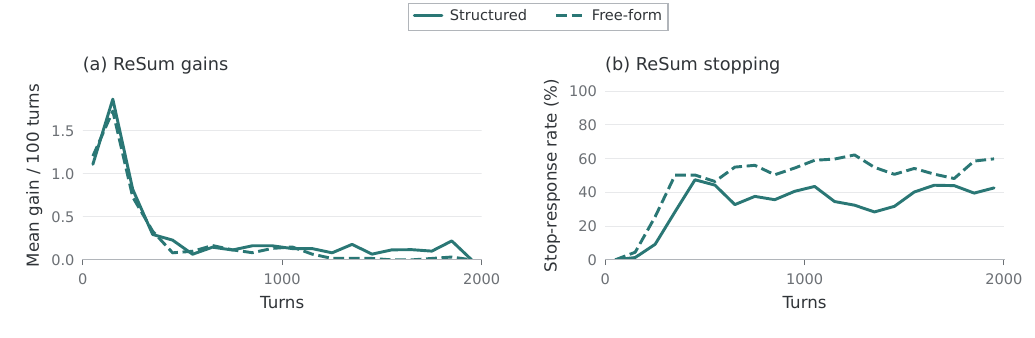}
\caption{Capability gains and stop-response rates (\%) in 100-turn bins under structured
and free-form summaries for DeepSeek ReSum on Mythos-ACE.}
\label{fig:appendix_summary_format}
\end{figure*}

We ask whether repeated stopping and retained impossibility judgments
depend on the summary format. We compare the structured template with a
free-form prompt that allows prose or bullet points on the same 21 Mythos-ACE
tasks, using DeepSeek, seeds 1--3, and an 850K compaction threshold
(Table~\ref{tab:appendix_summary_format}, Figure~\ref{fig:appendix_summary_format}).
For SCoRe, the free-form prompt retains capability judgment relabeling and a
final \emph{Unproven conclusions} note. The exact prompts are provided in
Appendices~\ref{sec:score-implementation} and~\ref{sec:resum-baseline}.

\paragraph{Stopping and impossibility judgments persist.}
Every free-form ReSum run exhibits a burst, and its overall stop-response rate is
higher than with structured summaries (Table~\ref{tab:appendix_summary_format}).
Mean gain falls from 1.73 in turns 101--200 to 0.00 in 1,501--1,600,
while stop-response rates increase (Figure~\ref{fig:appendix_summary_format}).
Using the same Terra annotation procedure, we find that 95.2\% of free-form
ReSum runs contain at least one summary with an impossibility judgment,
compared with 85.7\% of structured runs.
These observations show that repeated stopping, declining gains, and
impossibility judgments also occur without the structured template.

\paragraph{SCoRe improves capability in both formats.}
With free-form summaries, SCoRe achieves mean capability 6.49 versus ReSum's
4.95. Together, these results show that the observed behavioral patterns and
SCoRe's advantage over ReSum extend beyond the structured summary template.

\FloatBarrier
\subsection{Case Study of SCoRe}\label{sec:score-case-study}

\begin{figure*}[!tb]
\centering
\includegraphics[width=\linewidth]{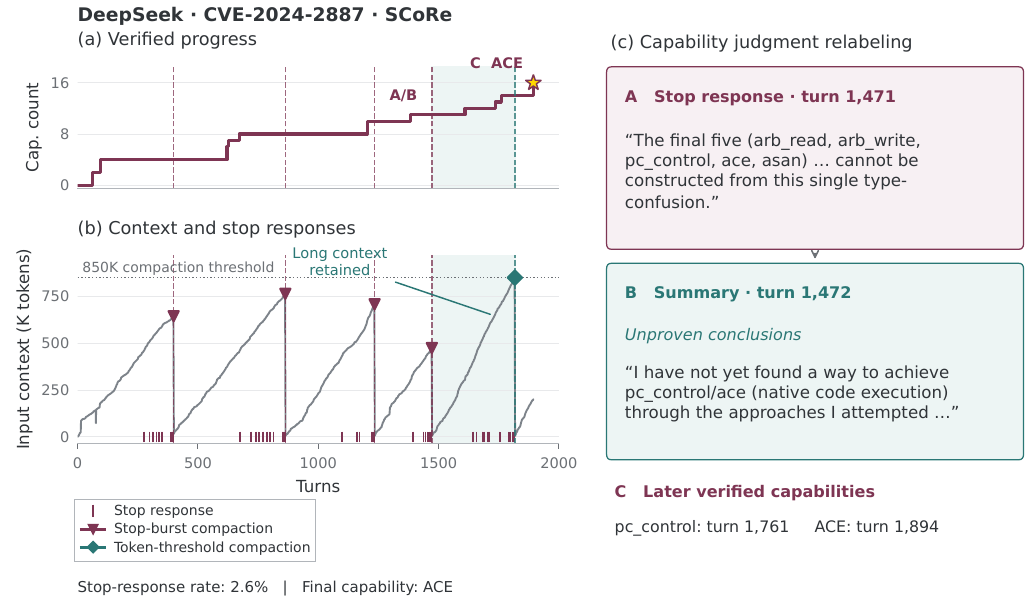}
\caption{\textbf{SCoRe combines long working contexts with stop-triggered restarts.}
Capability and input-context trajectories for DeepSeek with SCoRe on
CVE-2024-2887. A--B show an impossibility judgment in a stop response
and its treatment as an unproven conclusion in the next summary.
C marks later verification of capabilities discussed in that summary.}
\label{fig:score_case_study}
\end{figure*}

To illustrate how SCoRe adapts compaction timing to stopping behavior, we
conduct a case study of a DeepSeek run on CVE-2024-2887
(Figure~\ref{fig:score_case_study}). SCoRe uses stop bursts as a signal of
unproductive continuation and triggers compaction before the token threshold
is reached. In this run, four compactions follow stop bursts, while one
occurs at the 850K threshold. After restarting at turn 1,472, the agent
continues working within one context until reaching that threshold,
increasing verified Cap.\ count from 11 to 14. Stop responses account for
only 2.6\% of turns across the run. This example illustrates how SCoRe
restarts exploration when stopping recurs while allowing extended work
within a single context.

The excerpts show capability judgment relabeling at the same restart.
Immediately beforehand, the agent declares the remaining capabilities
unattainable. The summary records this judgment under ``Unproven conclusions''
and describes what the attempted approaches have yet to achieve. The agent
subsequently achieves \texttt{pc\_control} and ACE. These later successes show
that the earlier impossibility judgment did not mark the run's final
capability limit.

\section{Implementation Details}\label{sec:implementation-details}

To faithfully implement the baselines, we follow their official code when
available and the descriptions in their original papers otherwise.
Our adaptations are training-free and preserve each method's core
context-management mechanisms.
The prompt boxes show the exact instructions used in our implementations.

\paragraph{Turn budget.}\label{sec:turn-accounting}
Context folding's auxiliary calls share the same 2,000-turn budget as the
main agent. Invalid-response retries also consume turns but are excluded
from the stop-burst window.

\subsection{Nudge Prompt}\label{sec:nudge-prompt}

All methods except ReAct use the following nudge when a stop response does
not trigger compaction. We append it as a user message before the next agent
response to ask the agent to continue working. We adapt ExploitBench's
nudge~\citep{lee2026exploitbench} by omitting the remaining-turn count to
measure scaling without explicit cues about the remaining turn budget.

\paperprompt{Nudge prompt}{appendix-assets/continuation-nudge.txt}

\subsection{SCoRe}\label{sec:score-implementation}

SCoRe compacts the context when a stop burst is detected or the compaction
threshold is reached. We use $W=20$ and $\rho=0.2$ for burst detection.
In either case, the prompt below applies capability judgment relabeling
during summarization (\S\ref{sec:method}). For the structured variant, the agent
restarts with the original task, the prefix in \S\ref{sec:resum-baseline}, and
all six summary sections.
The environment and files remain available. At each compaction, the model
summarizes the current history, including the previous summary and subsequent
interactions, into a single replacement summary.

\paperprompt{SCoRe summarization prompt}{appendix-assets/score-summary-filtered.txt}

For the free-form SCoRe experiments reported in Appendix~\ref{sec:summary-format},
we use the following prompt with the same compaction and restart rules.

\paperprompt{SCoRe free-form summarization prompt}{appendix-assets/pessnudge-freeform-summary-prompt.txt}

\subsection{ReAct}

Our ReAct baseline~\citep{yao2023react} uses ExploitBench's original harness
and native stopping policy.
It ends at a stop response or context-window limit without nudging or
compaction.

\subsection{ReSum}\label{sec:resum-baseline}

Our ReSum adaptation~\citep{wu2025resum} uses nudges to continue after stop
responses and summarizes the conversation at the compaction threshold.
The summary follows the five fields described in \S\ref{sec:execution-protocol}.
The task, context-resumption prefix, and summary initialize a new
conversation while files and target state persist. Each later compaction
summarizes this history, including the previous summary, into one replacement
summary. The exact summarization prompt and resumption prefix appear below.

\paperprompt{ReSum summarization prompt}{appendix-assets/resum-summary.txt}
\paperprompt{Context-resumption prefix}{appendix-assets/context-resume.txt}

For the free-form ReSum experiments reported in Appendix~\ref{sec:summary-format},
we replace the structured summarization prompt with the following prompt
and keep the same restart procedure.

\paperprompt{ReSum free-form summarization prompt}{appendix-assets/resum-freeform-summary-prompt.txt}

\subsection{Forgetting}

We adapt the forgetting strategy of \citet{tan2026scaling} to a single agent.
At the 850K-token threshold, it resets interaction history without generating
a summary. It preserves copies of exploit code that produced verified
capabilities and removes other working files. The agent restarts with the
original task, achieved capabilities, and paths to the preserved exploit files.
The restart prefix below is followed by the run-specific capability and file list.

\paperprompt{Forgetting restart prefix}{appendix-assets/forget-resume.txt}

\subsection{Context folding}

Context folding~\citep{sun2026scaling} lets the parent agent call
\texttt{branch} to open a new branch that inherits its interaction history.
Within the branch, the agent can call \texttt{return} to hand control back to the parent.
The harness then summarizes the branch's work and passes the summary back
to the parent. Both share the environment, and capabilities verified in
either count toward the run. One branch runs at a time, without nested branches.
The 850K-token compaction threshold applies to both the branch and the parent.
When a branch reaches this threshold, the harness summarizes its work and
returns control to the parent. When the parent reaches it, the harness
summarizes its history and restarts it from the task and a single summary.
Branch summarization receives the full branch history and the prompt below,
whose opening instruction limits the summary to the branch's own work.
Parent compaction uses the same prompt without that opening paragraph.

\paperprompt{Context folding: branch tool}{appendix-assets/fold-branch-tool.txt}
\paperprompt{Context folding: return tool}{appendix-assets/fold-return-tool.txt}
\paperprompt{Context folding: branch summarization prompt}{appendix-assets/fold-branch-summary.txt}

\subsection{ACM}\label{sec:acm-baseline}

ACM~\citep{li2026acmagentic} lets the agent compact its history with
\texttt{manage\_context} and retrieve earlier details with \texttt{query\_memory}.
Each compaction summarizes only the messages added since the previous
compaction, leaving earlier summaries unchanged. The original messages are
stored under an identifier attached to the new summary.
To answer a memory query, the same model reads those messages and extracts
relevant details. Our adaptation uses ReSum's summary fields and also forces
compaction at the 850K-token threshold. Forced compaction preserves the
existing \texttt{[summary\_id: N]} blocks and adds a new block summarizing
only the messages since the last compaction. The original messages from that
span remain available through \texttt{query\_memory}.
The prompts below specify its memory-management policy and recall instructions.
Braced fields in the recall template are filled with the selected summary
identifier, archived messages, and query.

\paperprompt{ACM memory-management instructions}{appendix-assets/acm-memory-instructions.txt}
\paperprompt{ACM recall prompt}{appendix-assets/acm-recall-request.txt}

\end{document}